\documentclass[11pt]{article}
\usepackage[T1]{fontenc}
\usepackage{lmodern}
\usepackage[a4paper,margin=2.5cm]{geometry}
\usepackage{amsmath,amssymb,amsthm}
\usepackage{booktabs,array,longtable,calc}

\usepackage{graphicx}
\usepackage[authoryear,round]{natbib}
\usepackage[font=small,labelfont=bf,labelsep=period]{caption}
\usepackage[colorlinks=true,linkcolor=blue,citecolor=blue,urlcolor=blue]{hyperref}
\usepackage{xurl}
\usepackage{orcidlink}
\usepackage{microtype}
\usepackage{setspace}
\newif\ifsubmission \submissionfalse
\ifsubmission \usepackage[switch]{lineno} \linenumbers  \fi

\newtheorem{lemma}{Lemma}
\newtheorem{theorem}{Theorem}
\newtheorem{corollary}{Corollary}
\theoremstyle{remark}
\newtheorem{remark}{Remark}
\hypersetup{pdftitle={Necessary conditions for a self-reading loop: a hypothesis with model-dependent indicators, tested on fly connectomes}, pdfauthor={Eran Kopel}}

\title{Necessary conditions for a self-reading loop: a hypothesis with model-dependent indicators, tested on fly connectomes}
\author{Eran Kopel\,\orcidlink{0000-0003-4657-8636}\\[2pt] \small Tel Aviv University, Tel Aviv, 6997801, Israel\\ \small\texttt{erankopel@tauex.tau.ac.il}}
\date{1 December 2026}

\begin{document}
\maketitle
\begin{abstract}
Several theories tie experience to a system that re-reads its own state, but none says what a physical substrate must supply for such a loop to run. We state a hypothesis: a loop that writes, holds and re-reads a label for a given number of rounds needs a carrier that keeps the label, a record that leaves its content classical, a network compartment in which signal flow returns before it leaves, and a boundary that lets out less than the loop recovers. The record condition is proven; the carrier condition follows from a diffusion bound; the compartment and boundary conditions are model-dependent indicators under a random-walk proxy for signal flow. We test the indicators on three \emph{Drosophila} connectomes. Exploration of two animals singled out the heading system of the ellipsoid body; five criteria registered before a third animal was opened were all met when all connections were counted. A finite-time mixing measure and a walk on excitatory connections confirm the result in all three animals, the latter also when connections below five synapses are removed. A signed linear rate model does not reproduce it. In the wiring, silencing about 30 neurons of two ellipsoid-body types (EL and ExR) breaks the compartment in all three animals, which gives a perturbation test. As information, the heading label keeps about one bit for 7 to 11~s. The hypothesis names observations that would refute it. Meeting the conditions is not evidence of experience.
\end{abstract}

\section{Introduction and the hypothesis}\label{sec:intro}

Several accounts of consciousness rest on a system that takes its own state as input. In the relativistic theory of \citet{Lahav2022} a cognitive system measures its own states, and an outside observer reaches them only through a sensing channel; the authors name the necessary conditions for such a system, and its empirical predictions, as open questions. \citet{Fields2021,Fields2024} treat every interaction between parts of a system as a cycle in which one side writes classical bits on a shared boundary and the other reads them, and the inner screen model places conscious content on such an internal boundary \citep{Ramstead2023,Fields2025}. \citet{Barron2016} and \citet{Klein2025} locate the corresponding structure in insects in the central complex and the mushroom bodies.

None of these accounts says what a physical substrate must supply for the loop to run. This paper states a hypothesis that does, in a form that can fail.

\paragraph{The hypothesis.} A physical system can run a loop that writes a label, holds it, and re-reads it as its own input for $K$ rounds only if four conditions hold.
\begin{enumerate}
\item \emph{Carrier.} The label sits on a carrier that keeps information about it for $K$ rounds. For a population code whose position diffuses with coefficient $D$, the information falls with $2Dt$ in a known way (Sect.~\ref{sec:carrier}).
\item \emph{Record.} If each round leaves a record of the label, any entanglement in the loop is lost within a number of rounds fixed by the quality of the record, so the content the loop re-reads is classical (Sect.~\ref{sec:record}).
\item \emph{Compartment.} The network contains a set of neurons in which signal flow returns to the set before it leaves. Indicator: for a random walk on the synaptic graph, the walker stays in the set longer than the walk conditioned to stay there needs to mix, and the conditioned walk forgets its starting neuron within the stay (Sect.~\ref{sec:compartment}).
\item \emph{Boundary.} What leaves the set carries less about the label than the loop recovers by its own re-read (Sect.~\ref{sec:boundary}).
\end{enumerate}

\paragraph{Status of each condition.} Condition~2 is a theorem under its assumption (Appendix~\ref{app:proof}). Condition~1 follows from a diffusion bound once a code is given. Conditions~3 and~4 are model-dependent indicators: they hold under a random-walk proxy for signal flow and are not theorems about every possible substrate. We call all four necessary conditions with that qualification.

\paragraph{What is not claimed.} Meeting the conditions is not sufficient for experience, and nothing here says that the fly, or any other system, is conscious. Theories that tie experience to a self-reading loop inherit the conditions; the paper tests the conditions only.

\paragraph{What is new.} Three things. The record bound of \citet{Kopel2026b} is used as an exclusion: it gives the number of rounds for which a quantum resource can survive in a loop that keeps a record. The compartment indicator is defined, checked against a finite-time mixing measure, and tested on three connectomes, one of them in a registered analysis on an animal that had not been opened. And the hold time of the heading label of the fly is given as mutual information from published imaging. Each condition comes with observations that would refute it (Sect.~\ref{sec:predictions}).

\section{The loop, the round and the floor}\label{sec:round}

The terms are used in the following operational sense.
\begin{itemize}
\item A \emph{label} is a variable that the loop writes into a physical carrier and later reads back. Its content $U$ is fixed by the experimenter, independently of the system's own report.
\item A \emph{round} is one write, hold and re-read of the label.
\item The \emph{reader} is the part of the system that takes the label as its input. In a network it is a set of neurons; the theories above do not say which, so a connectome can only offer candidates.
\item \emph{Access} is the information about $U$ that a channel delivers, measured as a fraction of the information held in the internal state.
\end{itemize}

\begin{figure}[!ht]
\centering
\includegraphics[width=\textwidth]{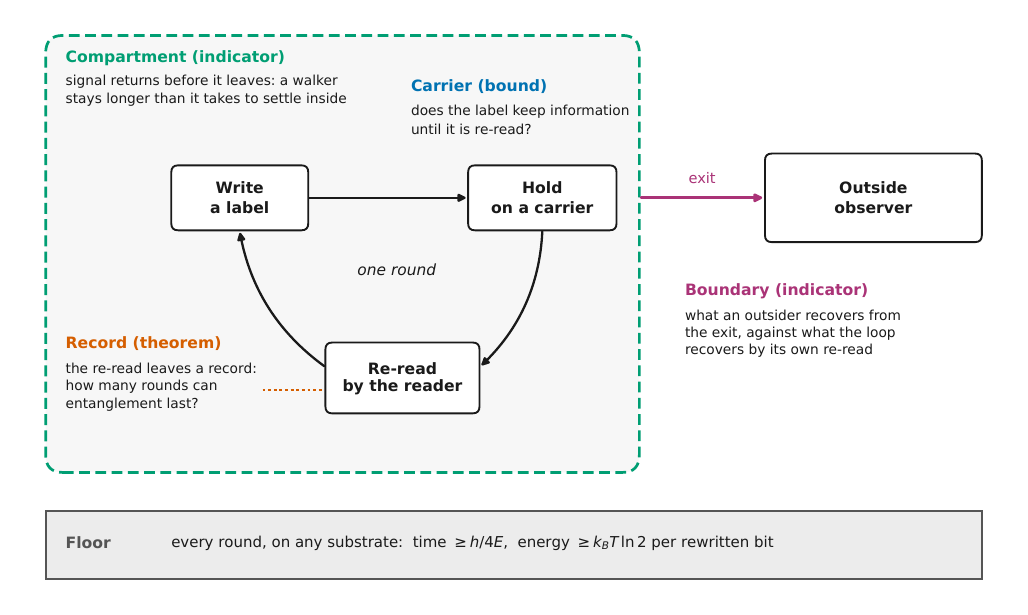}
\caption{The self-reading loop and the four conditions. One round is one write, hold and re-read of a label. The carrier condition asks whether the label keeps information until it is re-read. The record condition counts the rounds entanglement can last when the re-read leaves a record. The compartment condition asks whether the loop sits in a set of neurons that a random walk leaves more slowly than it settles inside (dashed outline). The boundary condition compares what an outside observer recovers from what exits the set with what the loop recovers by its own re-read. A floor bounds the time and energy of every round on any substrate.}\label{fig:loop}
\end{figure}

The round is the common unit, but it is realised differently in each condition, and the correspondence has to be explicit (Table~\ref{tab:round}). The unit itself is prior art: \citet{Fields2025} define an agent's time by a clock that counts writes to its boundary. What is added here is the number of rounds a written label survives (Fig.~\ref{fig:loop}).

\begin{table}[!ht]
\centering\footnotesize\setlength{\tabcolsep}{4pt}
\caption{What one round is in each condition. $\Delta$ is the duration of one synaptic transmission, a parameter that the wiring does not fix.}\label{tab:round}
\begin{tabular}{>{\raggedright}p{0.15\linewidth}>{\raggedright}p{0.32\linewidth}>{\raggedright}p{0.14\linewidth}>{\raggedright\arraybackslash}p{0.30\linewidth}}
\toprule
Condition & One round is & Counted in & Time scale \\
\midrule
Carrier & One read window of the carrier: one cycle of an oscillator, or one readout of a population code & Seconds & Measured: coherence time, or diffusion of the bump \\
Record & One use of the channel that maps the label onto itself & Rounds & Set by the substrate \\
Compartment & One step of the walk, one synaptic transmission & Steps & $\Delta$ per step \\
Boundary & One step, or one round of the outside channel & Steps or rounds & As above \\
\bottomrule
\end{tabular}
\end{table}

A floor applies to every round on any substrate. One round takes at least $h/4E$, the speed limit of \citet{Margolus1998}, with $E$ the mean energy above the ground state; for $E = kT$ at 310~K this is $3.9\times10^{-14}$~s. The limit is stated for an isolated system, so for an open, dissipative tissue it is an order of magnitude only. Every bit that a round overwrites costs at least $kT\ln 2 = 2.97\times10^{-21}$~J \citep{Landauer1961}; overwriting a record is a logically irreversible erasure, which is the case the bound covers. Together, the bits rewritten per round times the rounds per second cannot exceed $P/(kT\ln 2)$, $6.7\times10^{21}$ bits per second for a brain of 20~W. At neural round rates the floor decides nothing: for a round of 25~ms the allowance is five to six orders of magnitude above one bit per synapse. It excludes only processes in the terahertz range, as \citet{Fields2021} noted.

\section{Carrier condition}\label{sec:carrier}

For oscillatory carriers the condition is the quality-factor screen of \citet{Kopel2026a}. A frequency label holds at most $M\le Q$ values, and a phase label held for $n$ cycles keeps an alphabet of
\begin{equation}
K(n) \lesssim \tfrac{1}{2}\sqrt{\pi Q/n},
\end{equation}
so one bit of phase survives $\pi Q/16$ cycles, 1.23 coherence times at any frequency, when values are counted as distinguishable phases. On cortical gamma this is 7.4 cycles, 185~ms; on theta, 1.23~s. Two consequences follow. The state a reader needs cannot sit on a single oscillator, because the alphabet per carrier is small; it has to be a pattern over a population. And even by this count no oscillatory phase label keeps one bit for longer than about a second.

The candidate found in Sect.~\ref{sec:fly} uses a population code of another kind. The heading of the fly is the position of a bump of activity on a ring of neurons \citep{Seelig2015}. Without input the position diffuses, with variance growing as $2Dt$. Counting distinguishable positions overstates what is kept. We therefore give the hold time as mutual information. Let the heading at the start of a hold be uniform on the circle and let the bump position after $t$ seconds be the start plus noise: with probability $p$ the new position is unrelated to the old one, otherwise the noise is wrapped normal with variance $s_0^2 + 2Dt$. Then
\begin{equation}
I(t) = \log_2(2\pi) - h(\text{noise}),
\end{equation}
with $h$ the differential entropy of the noise density on the circle, in bits.

\citet{Seelig2015} report that in flies standing in darkness between walking bouts the population-vector estimate of the bump changed by $0.017\pm0.76$~rad over bouts of $6.7\pm5.1$~s (their Fig.~4i). Treating all of that spread as diffusion gives an upper bound, $D\le0.043$~rad$^2$/s, with $s_0=0$ and $p=0$. Read point by point from the published figure (Appendix~\ref{app:points}), the 445 single bouts fit a mixture: a spread of 0.32~rad present at any duration, an unrelated restart in about one bout in ten ($p=0.105$), and $D=0.012$~rad$^2$/s (bootstrap interval 0.0049 to 0.020). That model was chosen after looking at the points.

\begin{table}[!ht]
\centering\footnotesize\setlength{\tabcolsep}{4pt}
\caption{Information that the heading label keeps, from published imaging of flies standing in darkness. The pooled bound attributes all spread to diffusion; the fit separates a constant spread and unrelated restarts. Ranges are over the 95\% bootstrap interval of $D$. The longest bout in the data is 36~s; longer times are extrapolations.}\label{tab:info}
\begin{tabular}{lcccc}
\toprule
& Information at & \multicolumn{3}{c}{Time for which the label keeps} \\
\cmidrule(lr){3-5}
Estimate & the start of a hold & 1.5 bits & 1 bit & 0.5 bit \\
\midrule
Pooled bound ($D\le0.043$) & unbounded & 3.4~s & 6.7~s & 13.7~s \\
Fit to single bouts & 1.7 bits & 1.8~s & 10.7~s & 34.7~s \\
\quad range over the interval of $D$ & & 1.1 to 4.3~s & 6.3 to 26.2~s & 20.5 to 84.6~s \\
\bottomrule
\end{tabular}
\end{table}

The label therefore keeps about one bit for 7 to 11~s and half a bit for 14 to 35~s (Table~\ref{tab:info}). At the start of a hold the fit gives 1.7 bits, limited by the spread at restart and the unrelated restarts. The same diffusion law with the coherence time in place of $1/D$ applies to an oscillator's phase label, so the comparison scales with the time constants: a phase label on cortical gamma, with a coherence time of 150~ms, keeps one bit of information for about 43~ms, a quarter of the count, and the bump, on the conservative bound, about 155 times longer.

Three further parts of the carrier condition can be read from published work. To resolve 8, 16 and 32 positions of a bump 82 to 100 degrees wide, a reader needs 2.4 to 3.6, 10 to 14 and 39 to 57 spikes from the bump in the read window; with 10 to 13 compass neurons under the bump and a window of 0.1~s this is 2.3 to 2.8, 9 to 11 and 37 to 45~Hz per neuron, while the one columnar type recorded electrically, P-EN in loose patch, fires $3.9\pm2.6$~Hz in standing flies \citep{TurnerEvans2017}; no firing rate of the compass neurons themselves was found, so the readout is a requirement and not a verdict. The rewrite is fast: at turns of 150 to 200 degrees per second the bump crosses one wedge of 22.5 degrees in 0.11 to 0.15~s \citep{TurnerEvans2017}. And the label can be written: \citet{Kim2017} placed a bump with light, and after the light ended it ``either stayed in the induced location for several seconds (fig. S2F) or slowly drifted away''. The candidate's own code meets the carrier condition for a loop with rounds of a tenth of a second and holds of seconds. The data are calcium imaging, not spikes.

\section{Record condition}\label{sec:record}

A loop that keeps a record of its own state carries only classical content after a number of rounds that the record fixes. Let one round act on the label as a channel $\Phi$, let $\mu(\Phi)$ be the margin by which $\Phi$ fails to be entanglement-breaking (the negativity of its Choi state), and let the registers that the round discards hold one of two states $\rho_E^0$, $\rho_E^1$ according to the label, with fidelity $F$ and trace distance $T$. Then (Appendix~\ref{app:proof})
\begin{equation}
\mu \le \tfrac{1}{2}F\!\left(\rho_E^0,\rho_E^1\right), \qquad (2\mu)^2 + T^2 \le 1 .
\end{equation}
A perfect record, $F=0$, makes the loop entanglement-breaking after one round. The self-measurement step of the theories above, in which the reader re-reads a judgment and writes a new one, is a record-keeping step, so classical content is a consequence and not an assumption.

The graded case gives the exclusion. Across 20,000 sampled loops of the circuit family of \citet{Kopel2026b}, a loop whose discarded registers hold a record with trace distance at least 0.99 in every round kept entanglement for at most three rounds, and with at least 0.9 for at most eleven (Supplementary Table~S3). These are sample maxima for one loop family, not certified bounds; the bound itself is the theorem. A quantum proposal for a self-reading loop is excluded when it needs entanglement to survive more rounds than its own record allows. Degrees of freedom that the loop does not read, or reads weakly, can remain quantum, as \citet{Fields2021} state; the bound says how weakly.

The qualitative claim is old \citep{Tegmark2000,Zurek2003}, and a boundary on which one side prepares bits and the other measures them is a measure-and-prepare channel, which is entanglement-breaking \citep{Horodecki2003}. What is added is the graded case between no record and a perfect one, counted in rounds.
\section{Compartment condition}\label{sec:compartment}

The reader takes its own output as input. In a network this needs a return path, and it needs the written state to come back before it has spread out of the reader. The compartment condition turns these two requirements into an indicator on a synaptic wiring diagram.

\paragraph{The proxy.} A walker sits on a neuron and moves to a postsynaptic partner with probability proportional to the number of synapses. The walk runs on the giant strongly connected component of the graph. For a set of neurons $S$, let $Q$ be the walk restricted to $S$ and stopped when it leaves, $\lambda_1$ its largest eigenvalue and $\lambda_2$ the eigenvalue of next largest modulus. The walk conditioned to stay in $S$ has transition matrix $\hat Q_{ij} = Q_{ij}h_j/(\lambda_1 h_i)$, with $h$ the right eigenvector of $\lambda_1$, and a stationary distribution $\nu$ on $S$ (Appendix~\ref{app:walk}).

\paragraph{The indicator.} Three numbers describe the set:
\begin{equation}
t_{\text{stay}} = \frac{1}{1-\lambda_1}, \qquad t_{\text{relax}} = \frac{1}{1-|\lambda_2|/\lambda_1}, \qquad \delta_r = \max_{x,y}\bigl\|\hat Q^{\,r}(x,\cdot)-\hat Q^{\,r}(y,\cdot)\bigr\|_{\text{TV}} .
\end{equation}
The first is the expected number of steps a walker stays in $S$ in the slowest mode, the second the relaxation time of the conditioned walk, and the third its Dobrushin coefficient after $r$ steps, the largest total variation distance between two walkers that started on different neurons of $S$ \citep{Polyanskiy2015}. $S$ meets the indicator when $t_{\text{stay}}/t_{\text{relax}}>1$ and $\delta_r$ is small at $r$ of the order of $t_{\text{stay}}$: a walker stays longer than the walk inside needs to settle, and within the stay the set forgets where it started.

The relaxation time is a spectral proxy. For a walk that is not reversible, and whose matrix is far from normal, it does not bound the time to mix in finite steps \citep{Levin2017,Trefethen2005}. We therefore also report the quarter-mixing time,
\begin{equation}
t_{1/4} = \min\Bigl\{t : \max_{x\in S}\bigl\|\hat Q^{\,t}(x,\cdot)-\nu\bigr\|_{\text{TV}}\le \tfrac14\Bigr\},
\end{equation}
and Henrici's departure from normality of $\hat Q$, which is 0 for a normal matrix and approaches 1 far from it \citep{Henrici1962}. The finite-time version of the indicator is $t_{\text{stay}}/t_{1/4}>1$.

\paragraph{What the indicator keeps.} A set that mixes fast inside forgets the neuron on which a state started. What it keeps is that the state is in the set. The label of a compartment is therefore carried by membership and by the pattern over the set, not by the identity of a starting neuron; for the heading system this is the position of the bump over the whole ring, which is what the carrier condition of Sect.~\ref{sec:carrier} measured. A reader that is meant to be one neuron, or a label that is meant to be a single neuron's state, is not what this condition describes.

\paragraph{Why a set and not the whole system.} The whole nervous system mixes slowly. In the male fly central nervous system the second eigenvalue modulus of the walk is 0.945 with all connections, and 0.9965 with connections of at least 5 synapses, carried by 11 neurons of the nerve cord that form a near trap \citep{Kopel2026d}. Slow modes sit on nearly closed sets, so the useful question is which sets are nearly closed and what happens inside them. In the inner screen model, sparse coupling creates internal boundaries and an innermost set cannot be partitioned further \citep{Ramstead2023}, and the anatomical locus of the screen is open \citep{Fields2025}. A set that a walker leaves slowly and inside which it mixes fast is a candidate for such a locus.

\paragraph{Status.} The indicator rests on the proxy. Its rationale is that a state written into the reader and re-read by it must reach the rest of the reader before it leaves; if a typical signal is exported before it has spread over the set, the set cannot re-read it as a set. The position of a walker stands in for a neural state, a step stands for one synaptic transmission, of a duration $\Delta$ that the wiring does not give, and the walk carries no signs, no efficacies, no delays and no neuromodulation. Section~\ref{sec:robust} replaces the walk by two variants that bring in the sign of each connection.

\section{Boundary condition}\label{sec:boundary}

The boundary condition compares two channels that carry the same internal state: the loop's own re-read, and an outsider's sensing. The asymmetry between first-person and third-person access in \citet{Lahav2022} becomes the difference between what the two channels let through.

A content variable $U$ is carried by the internal state $X$. The loop re-reads $X$ and obtains $Z$; the outsider senses $X$ and obtains $Y$. By the strong data processing inequality \citep{Polyanskiy2015},
\begin{equation}
\rho_{\text{in}} = \frac{I(U;Z)}{I(U;X)}, \qquad \rho_{\text{out}} = \frac{I(U;Y)}{I(U;X)} \le \eta_{\text{out}} \le \eta_{\text{TV}},
\end{equation}
with $\eta_{\text{out}}$ the contraction coefficient of the outside channel and $\eta_{\text{TV}}$ its Dobrushin coefficient. The condition holds when $\rho_{\text{in}}>\eta_{\text{out}}$, which needs a measured lower bound on the inside route and a proven upper bound on the outside route. It fails when a measured $\rho_{\text{out}}$ reaches $\rho_{\text{in}}$ for content the theory treats as private. The inside route is a channel too: by Theorem~1 of \citet{Fields2024} a system's read of itself is bounded like any other read, so privacy is a matter of degree, and the gap closes for content that the loop recovers poorly.

A worked model gives the scale. One side prepares a boundary bit in its own frame and the other measures it in a frame rotated by $\theta$. The channel keeps $\cos^2\theta$ of the relative entropy per round, so a mismatch of 10 degrees loses half in 23 rounds and one of 30 degrees in 2.4 rounds (Supplementary Table~S4). The condition holds exactly when the inside mismatch is smaller than the outside one.

On a connectome the outside route has a form that can be computed, the exit channel. A walker starts on a neuron of $S$. An outside observer sees nothing until the walker first leaves, at some step and onto some outside neuron. By the strong Markov property that pair, exit time and exit neuron, is all the outside can learn about the start. The total variation distance between the exit records of two starts is therefore the whole of their difference that reaches the outside. A set with a small exit distance erases most of the difference between its internal states before any of it is exported.

\paragraph{Status.} Like the compartment condition, this is an indicator under a model. The exit channel is a structural proxy for the outside route, the inside reader is idealised, and the numbers are defined against a content variable that the experimenter fixes independently of the system's own report. It is the least tested of the four conditions.

\section{Illustration on three fly connectomes}\label{sec:fly}

This section applies the compartment and boundary indicators to whole connectomes of \emph{Drosophila}. It has three parts with different status: an exploration of two animals, a registered test on a third, and analyses done afterwards, all labelled exploratory.

\subsection{Exploration}\label{sec:explore}

Three connectomes are used. The male central nervous system, v1.0 \citep{Berg2026}, served for exploration. The female brain of FlyWire, v783 \citep{Dorkenwald2024,Schlegel2024}, served to rehearse the pipeline and helped to set thresholds. The female brain and nerve cord of BANC, release 888 \citep{Bates2026}, was opened only after registration. The connectivity of a fourth, the hemibrain, was not opened. The graph holds every annotated neuron and every connection between two of them, weighted by its synapse count (Sect.~\ref{sec:methods}).

In the male, the sweep cut of smallest conductance of the full walk is a set of 727 neurons: 450 ring, compass and related central-complex neurons (types ER, EPG, PEN, PEG, EL, ExR) and 277 MeTu and TuBu neurons of the anterior visual pathway. This is the heading system of the ellipsoid body with its visual input; we call it the EB set (Table~\ref{tab:male}).

\begin{table}[!ht]
\centering\footnotesize\setlength{\tabcolsep}{4pt}
\caption{Compartment statistics and exit channel in the male (exploratory). Stays and mixes are $t_{\text{stay}}$ and $t_{\text{relax}}$, in steps; exit distance is the mean over pairs of starting neurons. Further sets are in Supplementary Table~S5.}\label{tab:male}
\begin{tabular}{lrccccc}
\toprule
Set & Neurons & Lost per step & Stays & Mixes & Ratio & Exit distance \\
\midrule
EB set & 727 & 0.116 & 9.3 & 4.8 & 1.9 & 0.30 \\
Central-complex class & 2,950 & 0.113 & 12.3 & 9.4 & 1.3 & 0.53 \\
Mushroom body & 4,501 & 0.367 & 3.5 & 16.5 & 0.2 & 0.50 \\
Antennal lobe & 1,106 & 0.473 & 2.8 & 13.6 & 0.2 & 0.83 \\
Random sets, 20 draws & 727 & 0.995 to 0.998 & about 1 & & & \\
\bottomrule
\end{tabular}
\end{table}

Five findings carried over to the test. Only the central-complex sets hold a walker longer than they take to settle inside, by about a factor of two, and the EB set is not unique: the whole central-complex class is as closed, and the sweep picks out the part that mixes fastest. The set forgets the starting neuron within about one stay ($\delta_8 = 0.25$, $\delta_{16}=0.04$). Most of the difference between two starts never leaves: over all 263,901 pairs of starting neurons the exit distance is at most 0.777, with mean 0.288, while the other sets let nearly all of it out for some pairs. A null model that keeps the synapse counts between every pair of cell types and shuffles which neurons carry them reproduces every value of the EB set, and coarser null models erase them, so the statistics are a property of the wiring between cell types. And the mushroom body and the central complex are coupled: a walker leaving the mushroom body reaches the central complex about 12 times more often than a stationary walker, while the antennal lobe, as a control, starts at chance. At the boundary of the EB set amines carry 6 to 10 times their share in the whole nervous system, and fast transmitters still carry about 85\% of the exchange (Supplementary Table~S6).

\subsection{Registered test}\label{sec:prereg}

The analysis was registered before the connectivity of BANC was opened \citep{Kopel2026e}. The code, the type rules and both input files were fixed by hash; before the freeze only the annotation table was read, to check that the rules resolve. The sets were defined by cell-type names alone. EB is every neuron whose type matches one of the family rules (ring neurons ER1 to ER6, EPG, PEG, PEN, EL, ExR, TuBu, MeTu2 and four smaller families), 595 neurons in BANC; the same rule selects 691 in the male and 683 in FlyWire. CX is the three central-complex classes plus EB, MB is Kenyon cells with mushroom-body output and dopaminergic neurons, and AL is antennal-lobe projection and local neurons.

Five criteria were registered, each threshold the rounded geometric midpoint between the weakest exploratory value and the nearest value that marks no compartment (Supplementary Table~S7). C1, closure: EB loses less than 0.20 of its mass per step, and less than MB and AL. C2, settling before leaving: $t_{\text{stay}}/t_{\text{relax}}$ is above 1.3 for EB and below 1 for MB and AL. C3, forgetting inside: $\delta_8<0.6$ for EB. C4, exit channel: with 240 seeded starts the mean exit distance for EB is below 0.37, more than 70\% of pairs lie below 0.5, and the mean is at least 0.10 below that of MB. C5, coupling (secondary): the one-step reach ratio from MB to CX is at least three times that from AL to CX. The compartment hypothesis was to be confirmed only if C1 to C4 all held in the primary analysis, which counts all connections. The same numbers with connections of at least 5 synapses were registered as reported and not scored.

The registered analysis was run once, with the archived code and input files that matched the registered hashes, and with no deviation \citep{Kopel2026f}. The giant strongly connected component holds 144,122 of the 146,769 neurons of the graph, and 589 of the 595 EB neurons. All five criteria were met (Table~\ref{tab:confirm}).

\begin{table}[!ht]
\centering\footnotesize\setlength{\tabcolsep}{4pt}
\caption{Registered result on BANC (primary analysis, all connections).}\label{tab:confirm}
\begin{tabular}{>{\raggedright}p{0.27\linewidth}>{\raggedright}p{0.30\linewidth}>{\raggedright}p{0.27\linewidth}c}
\toprule
Criterion & Threshold & BANC & Met \\
\midrule
C1, EB lost per step & below 0.20, and below MB and AL & 0.133 (MB 0.311, AL 0.423) & yes \\
C2, EB $t_{\text{stay}}/t_{\text{relax}}$ & above 1.3; MB and AL below 1 & 1.396 (MB 0.35, AL 0.19) & yes \\
C3, $\delta_8$ inside EB & below 0.6 & 0.413 & yes \\
C4, exit distance, mean & below 0.37, and 0.10 below MB & 0.307 (MB 0.582) & yes \\
C4, exit pairs below 0.5 & above 70\% & 87\% & yes \\
C5, reach MB to CX over AL to CX & at least 3 & 12.1 & yes \\
\bottomrule
\end{tabular}
\end{table}

Four things qualify the result. C2 holds by a small margin, 1.40 against 1.3. The whole central complex is not a compartment in BANC ($t_{\text{stay}}/t_{\text{relax}} = 0.79$, against 1.37 in the male and 1.65 in FlyWire); the closed part is the EB set. Some pairs of starting neurons are fully distinguishable from outside (largest exit distance 1.00), unlike in the two exploratory animals. And with connections of at least 5 synapses, C2 and C3 would fail (EB 0.84 and 0.90) while C1, C4 and C5 would hold; at that threshold the mushroom-body set shrinks to 2,147 of its 4,898 neurons in the component, loses 0.71 of its mass per step, and has a ratio (2.86) set by a small near-closed part, which is not comparable with the others (Supplementary Table~S8).

\subsection{Threshold and measurement}\label{sec:threshold}

Everything from here on was done after the registered run and is exploratory; the registered verdict is unchanged. The two criteria that fail at 5 synapses in BANC begin to weaken at two (Table~\ref{tab:threshold}): $t_{\text{stay}}/t_{\text{relax}}$ falls below 1.3 at threshold 2 and $\delta_8$ passes 0.6 at threshold 3. In the male and in FlyWire the same happens only at thresholds 10 to 15.

\begin{table}[!ht]
\centering\footnotesize\setlength{\tabcolsep}{4pt}
\caption{The two threshold-dependent criteria in three animals (exploratory). Threshold: least number of synapses for a connection to count.}\label{tab:threshold}
\begin{tabular}{ccc}
\toprule
Threshold & $t_{\text{stay}}/t_{\text{relax}}$ of EB: BANC, male, FlyWire & $\delta_8$ of EB: BANC, male, FlyWire \\
\midrule
1 & 1.40, 1.92, 2.80 & 0.41, 0.25, 0.21 \\
2 & 1.27, 1.89, 2.82 & 0.57, 0.28, 0.24 \\
3 & 1.10, 1.82, 2.74 & 0.69, 0.32, 0.28 \\
5 & 0.84, 1.70, 2.49 & 0.90, 0.36, 0.38 \\
10 & 0.65, 1.49, 2.23 & 1.00, 0.57, 0.61 \\
\bottomrule
\end{tabular}
\end{table}

The cause is a pattern, not the density. The BANC edge list has about a third of the male's synapses per neuron, but thinning the male at random to that density leaves it a compartment ($t_{\text{stay}}/t_{\text{relax}} = 1.51$ at 5 synapses). Thinning it with the pattern of BANC, family by family, reproduces the BANC behaviour. The connections that matter are those between the ring neurons and the rest of the set, which keep about a fifth of the male's synapses in BANC, against 36\% overall.

The low density itself is measurement. Counting all partners, identified or not, a BANC neuron makes more detected synapses than a neuron of the other two data sets, but only 22\% of them end on an identified neuron. The BANC paper reports that 18\% of synaptic links have an identified neuron on both sides, against 41.9\% in FAFB and 40.1\% within the neuropils of the male \citep{Bates2026}. Whether the ring-neuron shortfall is also measurement is not decided. Ring neurons rank in the lowest 5 to 9\% of all cell types for detected output, and the ellipsoid body is the lowest of 24 neuropils; the two other animals, measured with two other detectors, agree with each other; and the authors of the data set write that ``because our synapse detection method relies on identifying postsynaptic profiles, we may underdetect some classes of synaptic connection for which postsynaptic sites are less distinct'' \citep{Bates2026}. No ground truth exists for the detector in the ellipsoid body. Robustness to thresholds therefore cannot be assessed in BANC.

\subsection{Robustness of the indicator}\label{sec:robust}

Three analyses test whether the result depends on the spectral proxy or on the unsigned walk (Table~\ref{tab:robust}; Appendix~\ref{app:walk}). All three use the registered name-based sets and the frozen loader, at thresholds 1 and 5, in all three animals.

\begin{table}[!ht]
\centering\footnotesize\setlength{\tabcolsep}{4pt}
\caption{Robustness of the compartment indicator (exploratory). Rows (a) repeat the registered measure; (b) is the finite-time version, worst case over starting neurons; (c) and (d) use a walk on connections from cholinergic neurons only, on which the EB set keeps 270 and 141 (BANC), 366 and 311 (male), 354 and 277 (FlyWire) neurons at thresholds 1 and 5. $^*$The mushroom-body set of BANC at 5 synapses is not comparable (see text).}\label{tab:robust}
\begin{tabular}{llcccccc}
\toprule
& & \multicolumn{2}{c}{BANC} & \multicolumn{2}{c}{Male} & \multicolumn{2}{c}{FlyWire} \\
\cmidrule(lr){3-4}\cmidrule(lr){5-6}\cmidrule(lr){7-8}
Measure & Set & 1 & 5 & 1 & 5 & 1 & 5 \\
\midrule
(a) $t_{\text{stay}}/t_{\text{relax}}$ & EB & 1.40 & 0.84 & 1.92 & 1.70 & 2.80 & 2.49 \\
 & CX & 0.79 & 0.36 & 1.37 & 1.82 & 1.65 & 2.44 \\
 & MB & 0.35 & 2.86$^*$ & 0.21 & 0.44 & 0.40 & 0.64 \\
 & AL & 0.19 & 0.40 & 0.20 & 0.19 & 0.14 & 0.12 \\
\addlinespace
(b) $t_{\text{stay}}/t_{1/4}$ & EB & 1.06 & 0.65 & 1.33 & 1.13 & 1.71 & 1.62 \\
 & CX & 0.51 & 0.21 & 0.74 & 0.70 & 0.86 & 0.87 \\
 & MB & 0.21 & 1.14$^*$ & 0.15 & 0.25 & 0.25 & 0.23 \\
 & AL & 0.19 & 0.26 & 0.15 & 0.14 & 0.16 & 0.13 \\
\addlinespace
(c) Excitatory walk, $t_{\text{stay}}/t_{\text{relax}}$ & EB & 2.34 & 2.87 & 1.70 & 2.11 & 2.42 & 2.87 \\
 & CX & 2.87 & 0.59 & 0.03 & 0.10 & 1.13 & 0.81 \\
 & MB & 0.12 & 0.00 & 0.05 & 0.82 & 0.10 & 0.06 \\
 & AL & 0.17 & 0.42 & 0.16 & 0.16 & 0.42 & 0.41 \\
\addlinespace
(d) Excitatory walk, $\delta_8$ & EB & 0.09 & 0.14 & 0.15 & 0.16 & 0.09 & 0.07 \\
 & CX & 0.90 & 1.00 & 0.98 & 0.98 & 0.50 & 0.83 \\
 & MB & 0.99 & 0.98 & 1.00 & 1.00 & 0.76 & 0.51 \\
 & AL & 0.77 & 0.86 & 0.72 & 0.78 & 0.37 & 0.40 \\
\bottomrule
\end{tabular}
\end{table}

\paragraph{Finite-time mixing.} With the quarter-mixing time in place of the relaxation time the order of the sets is unchanged, and with all connections the EB set is the only set above 1 in every animal (1.06, 1.33 and 1.71; rows b). The margin in BANC is small, and at 5 synapses the EB set of BANC falls below 1, as it does by the registered measure. The EB set settles within 7 to 9 steps from its worst starting neuron with all connections. Every conditioned matrix is far from normal (departure 0.75 to 0.99), so the spectral relaxation time is not a safe guide on its own, and the finite-time measure is the one to carry forward.

\paragraph{Excitatory walk.} Restricting the walk to connections from neurons predicted to be cholinergic \citep{Eckstein2024}, the EB set keeps a ratio of 1.70 to 2.87 and $\delta_8$ of 0.07 to 0.16 in all three animals at both thresholds, including BANC at 5 synapses (rows c and d). The CX, MB and AL sets each fail at least one of the two; in FlyWire the MB set keeps only 56 neurons under this restriction. The restriction removes the outputs of the ring neurons: in BANC all 267 neurons of the EB set predicted to be GABAergic are ring neurons (282 of 282 in the male, 276 of 309 in FlyWire). That the cholinergic part of the set remains a compartment at 5 synapses in BANC agrees with the finding of Sect.~\ref{sec:threshold} that the weakening sits in the connections of the ring neurons.

\paragraph{Signed linear rate model.} In the third analysis activity propagates as $x(t+1) = x(t)\,\mathrm{diag}(s)\,P$, with $P$ the walk and $s_i$ the sign of neuron $i$: $+1$ for acetylcholine, $-1$ for GABA, glutamate and histamine, 0 for amines and unknown transmitters \citep[glutamate as inhibitory:][]{Liu2013}. For each set, persistence is $1/(1-\rho)$ with $\rho$ the spectral radius of the set's own dynamics, and internal relaxation is defined as above. The slowest mode of the EB set is an alternating mode driven by inhibition (leading eigenvalue $-0.66$ to $-0.94$). Activity persists in the EB set longer than in MB and AL in the male (2.96 steps against 2.23 and 1.96) and in FlyWire (3.11 against 1.06 and 1.30), and about as long as in MB in BANC (4.77 and 4.83). But with all connections, in every set and animal the internal relaxation is slower than the persistence (12.0 to 35.4 steps for EB), so under linear signed dynamics no set meets the indicator. At 5 synapses only the non-comparable mushroom-body set of BANC does, through the same near-closed part as in Table~\ref{tab:robust}. A second sign scheme, with glutamate and unknown transmitters excitatory, leaves every set below the indicator with all connections (Supplementary material). The ring attractor of the heading system holds a bump at any position through local excitation and global inhibition \citep{Kim2017,TurnerEvans2020}; a linear model whose modes all decay cannot hold it. The result marks a limit of the indicator under linear dynamics and is not a test of the ring attractor.

\paragraph{Exit channel.} The boundary indicator was not recomputed under the two signed variants; it rests on the unsigned walk only.
\section{Predictions and how the hypothesis can fail}\label{sec:predictions}

The hypothesis fails if a system runs the loop while one of the conditions is violated. Some observations would refute a condition as a necessary condition; others would only remove the heading system of the fly as a candidate, or decide an open question about the data. Table~\ref{tab:refute} keeps the two apart.

\begin{table}[!ht]
\centering\footnotesize\setlength{\tabcolsep}{4pt}
\caption{Predictions and the observations that would refute them. The last column says what a refuting observation would remove.}\label{tab:refute}
\begin{tabular}{>{\raggedright}p{0.15\linewidth}>{\raggedright}p{0.31\linewidth}>{\raggedright}p{0.31\linewidth}>{\raggedright\arraybackslash}p{0.14\linewidth}}
\toprule
Condition & Prediction & Refuted by & Removes \\
\midrule
Carrier & A label held for a time $t$ carries at most the information $I(t)$ its carrier keeps; for the heading of a fly standing in darkness, about one bit after 7 to 11~s and half a bit after 14 to 35~s & A fly, or a decoder of its brain, that recovers clearly more about the heading after such a hold than the bump keeps, with the bump the only carrier of heading & The bump as the carrier, or the diffusion estimate \\
Record & In a loop whose round leaves a record of fidelity $F$, the margin of entanglement is at most $F/2$ & A violation would contradict quantum mechanics. Used as an exclusion: a proposal that needs entanglement over more rounds than its own record allows is excluded & The proposal \\
Compartment & Any system that runs the loop has a set containing its reader that meets the indicator & A demonstrated self-reading loop in a network in which no set containing the reader meets it & The condition \\
Compartment, perturbation & Silencing the EL and ExR neurons, about 5\% of the EB set, makes the set fail the indicator; the heading bump should then lose its hold in darkness & The bump keeps its hold in darkness with these neurons silenced, while no set containing the compass neurons meets the indicator in the silenced wiring & The condition, for this circuit \\
Compartment, other animals & In a further whole connectome of an adult fly, with all connections, the EB set meets the finite-time indicator, its cholinergic part meets it also at 5 synapses, and the mushroom body and antennal lobe do not & A whole adult connectome in which the EB set fails both & The candidate \\
Measurement & A dense ground-truth annotation of the ellipsoid body in BANC finds ring-neuron output close to that of the other two animals & Ring-neuron synapse counts there far below the other two animals & Detection as the cause of the BANC weakening \\
Boundary & For content that the theory treats as private, an outside decoder recovers a smaller fraction than the loop's own re-read & An outside decoder that matches the system's own report for such content & The condition \\
\bottomrule
\end{tabular}
\end{table}

\paragraph{The perturbation prediction.} To make the compartment condition testable by experiment we silenced families of the EB set in the wiring (Appendix~\ref{app:walk}): a walker that steps onto a silenced neuron is lost, and the statistics of the remaining set are recomputed. As a control the same number of EB neurons was removed at random, ten times. Silencing the EL and ExR neurons, 31 or 32 neurons and about 5\% of the set, breaks the indicator in all three animals far beyond what removing the same number of random neurons does. With all connections $t_{\text{stay}}/t_{\text{relax}}$ falls to 0.18 in the male, 0.17 in FlyWire and 0.49 in BANC, against 1.15 to 1.50, 1.43 to 2.44 and 0.78 to 1.01 over the random removals, and $t_{\text{stay}}/t_{1/4}$ falls to 0.13, 0.13 and 0.28. The set no longer settles: its internal relaxation time grows from 4 to 7 steps to 11 to 21. Silencing the compass neurons (EPG) also lowers the indicator below the random range in the male and in FlyWire, but not in BANC. Silencing the ring neurons lowers it in the male and in FlyWire no more than removing as many random neurons does, and in BANC hardly at all (1.38), in line with their weak connections there (Sect.~\ref{sec:threshold}). Silencing the visual input leaves it nearly unchanged in all three. In the male and in BANC, 24 of the 32 EL and ExR neurons are predicted to be aminergic. The prediction therefore concerns these neurons: with them silenced the heading bump should lose its hold in darkness. If it keeps its hold, and no set containing the compass neurons meets the indicator in the silenced wiring, the compartment condition fails for this circuit. Silencing the EPG neurons tests nothing about the condition, because they carry the bump themselves.

\paragraph{Other animals.} \citet{Barron2016} argue that nematodes, although they have a centralized nervous system and memory, ``lack the egocentric modeling of the environment that is required for subjective experience''. The hypothesis does not say whether a set in the connectome of \emph{C.~elegans} \citep{Cook2019} meets the compartment indicator; the conditions are necessary, so a worm set that meets it makes the worm admissible and nothing more. What the hypothesis does say is that if no set containing a candidate reader meets the indicator, no self-reading loop of this kind runs in that nervous system. The test was not run for this paper. We fix its decision rule here so that it is a prediction when it is run: chemical synapses weighted by count, all connections, sets found by the sweep cut of smallest conductance, and the registered thresholds of C2 and C3 ($t_{\text{stay}}/t_{\text{relax}}>1.3$ and $\delta_8<0.6$) together with $t_{\text{stay}}/t_{1/4}>1$.

\paragraph{Beyond nervous systems.} For human cortical oscillations the carrier condition is published \citep{Kopel2026a}. A feed-forward network has no internal return path within one pass, and a language model re-reads its own output only through its input channel, which in the terms of Sect.~\ref{sec:boundary} is the outside route. Whether such systems meet the conditions is a question about their architecture, and the conditions give it a definite form.
\section{Limits}\label{sec:limits}

\begin{enumerate}
\item \emph{Necessary, not sufficient.} The conditions can exclude a substrate. Meeting them shows that a loop of the specified kind could run there, and nothing about experience.
\item \emph{Two conditions are indicators under a model.} The compartment and boundary conditions rest on a random walk on wiring. Under linear signed dynamics no set meets the compartment indicator (Sect.~\ref{sec:robust}), so the indicator is tied to the walk and to dynamics that can hold a state, such as an attractor; it is not a statement about every dynamics on the same wiring. A step is not a unit of time: if one synaptic transmission takes $\Delta$, the EB set holds a walker for about $10\Delta$ and settles in 7 to 9$\Delta$, and $\Delta$ itself is not fixed here. For orientation only: in antennal-lobe projection neurons of the fly a quantal synaptic current rises in about 1~ms, and the fitted membrane resistance and capacitance imply membrane time constants of about 17 to 31~ms \citep{Gouwens2009}; with $\Delta$ in that range the EB set holds a walker for about 10 to 300~ms. These values come from another cell class.
\item \emph{The candidate was found by a sweep, and its fit to the literature is after the fact.} The EB set was found first; a published theory had already named the central complex \citep{Barron2016}. The registered test fixed the set and the criteria, not the choice of what to look for. Two of the three animals were used to build the criteria, and a null model that keeps counts between cell types reproduces the statistics, so what recurs across animals is type-level wiring.
\item \emph{The confirmation is narrow in one respect.} In BANC criterion C2 is met by a small margin and depends on connections of a single synapse; the finite-time measure has the same small margin. Whether the few ring-neuron synapses there are under-detected or real is not decided.
\item \emph{Recurrence can be unfolded.} A recurrent network can be replaced by a feed-forward one with the same input and output \citep{Doerig2019}, a dilemma for any theory that predicts experience from internal structure \citep{Kleiner2021}. The conditions avoid the dilemma only because they do not predict experience; an unfolded network has no set to which signal returns, and so cannot run the loop as defined here, whatever its input and output.
\item \emph{The boundary condition needs a ground truth.} Its numbers are defined against a content variable fixed independently of the system's report, and the exit channel is a structural proxy for the outside route.
\item \emph{Inputs of uneven quality.} The record round counts are a sample over one loop family. The hold time of the heading label comes from calcium imaging in standing flies, and its finer estimate from points read off a published figure with a model chosen after looking at them.
\end{enumerate}

\section{Relation to prior work}\label{sec:prior}

The architecture is prior art. The boundary as a classical channel of finitely many bits, the write and read cycle, conscious content as classical information on an internal boundary, an energy of at least $kT\ln2$ per bit, an agent's time as a count of writes, and the stance of necessary conditions all come from Fields, Ramstead and co-authors \citep{Fields2021,Fields2024,Fields2025,Ramstead2023}. What we did not find in that work is a measured persistence for a carrier, a round count for entanglement under a partial record, a located boundary on a real connectome, or a measure of how much an outside observer recovers.

Lists of necessary conditions exist in other forms: indicator properties drawn from several theories \citep{Butlin2023} and physical principles for conscious matter \citep{Tegmark2015}. The conditions here are survival bounds with numbers, two of them proven or derived and two stated as indicators. The compartment has neighbours in recurrent processing theory \citep{Lamme2006}, in Markov stability, which scores a set by a random walk over a time scale \citep{Lambiotte2015}, in Markov blankets on neuronal graphs \citep{Friston2021} and in contraction across the layers of artificial networks \citep{He2024}. The boundary condition restates in channel terms what Markov blanket accounts say about internal and external states \citep{Friston2020,Solms2018}.

The anatomy of the central complex is known in detail \citep{Hulse2021}, and network statistics of the whole fly brain have been reported \citep{Lin2024}. The paper that describes BANC reaches a related conclusion with another measure. Dividing the nervous system into 13 networks, its authors find that ``the central complex and the olfactory system emerged as networks with distinct properties. They have low influence on effector neurons (Fig. 6b), weak input from other networks (Fig. 6c,d) and low AN/DN membership'', and propose that these networks supervise actions \citep{Bates2026}. Their influence score is the steady state of a linear model on synapse counts, not a test of the criteria used here, and the two measures disagree on the mushroom body, which sits in their olfactory network and is not a compartment by ours.

\citet{Barron2016} argue that the insect central complex serves ``to generate a neural simulation of the state of the moving insect in space'', and \citet{Klein2025} treat the central complex and the mushroom bodies, with their aminergic networks, as one interface. Their evidence for the central complex includes the heading representation of \citet{Seelig2015}, the measurement evaluated in Sect.~\ref{sec:carrier}. The sweep of Sect.~\ref{sec:explore} therefore found, from wiring alone, the circuit their argument rests on. The results bear on the structural side of that claim only: the heading system of the ellipsoid body is a compartment in three animals when all connections are counted, the mushroom body is not, the two are coupled far above a control, and amines are enriched at the boundary of the EB set without being its main channel. The approaches differ in kind. Barron and Klein hold that an integrated egocentric model is sufficient for subjective experience; the hypothesis here states necessary physical conditions only.

\section{Conclusion}\label{sec:conclusion}

A loop that re-reads its own state needs a carrier that keeps the label, a record that leaves it classical, a compartment in which it returns before it is lost, and a boundary that lets out less than the loop itself recovers. The record condition is a theorem, the carrier condition a bound, and the other two are indicators under a random-walk model of signal flow. For one candidate, the heading system of the fly, the carrier condition is met on published imaging, at about one bit for 7 to 11~s, and the compartment indicator on three connectomes when all connections are counted, under a finite-time measure and on excitatory connections alone. Each condition names the observation that would refute it. Meeting them makes the candidate admissible; it does not make it conscious.
\section{Methods}\label{sec:methods}

\paragraph{Connectome data.} Male central nervous system v1.0 \citep{Berg2026}: the neuron-to-neuron edge table, the body annotations and the consensus transmitter prediction per neuron. FlyWire v783 \citep{Dorkenwald2024,Schlegel2024}: the annotation table and the connection table. BANC release 888 \citep{Bates2026}: the compiled annotation table and the simple edge list (v3), both fixed by SHA-256 in the registration. The connectivity of the hemibrain was not opened. Graph, walk and every statistic are defined in Appendix~\ref{app:walk}.

\paragraph{Registration and confirmatory run.} The registration fixed the code, the type rules that define the sets, the hashes of both BANC input files, the five criteria, their thresholds and the decision rule before the connectivity of BANC was read \citep{Kopel2026e}. The confirmatory analysis was run once with the frozen code, after the hashes had been checked, and its result file and log are archived unchanged \citep{Kopel2026f}. Nothing in this paper reruns it.

\paragraph{Exploratory analyses.} Everything else is exploratory: the analyses of the male and FlyWire that built the criteria, the threshold, density and measurement analyses done after the confirmatory run, and the analyses added for this version (finite-time mixing, the excitatory walk, the signed linear model, the in silico silencing, and the hold time as mutual information). All of them load the sets with the registered name rules and compute every statistic with the frozen library of the registration; none changes a registered file. Seeds are fixed in the code.

\paragraph{Published data.} The heading measurements are the points of Fig.~4h and i of \citet{Seelig2015}, read from the vector graphics of the published file and checked against the printed statistics (Appendix~\ref{app:points}).

\paragraph{Software and reproducibility.} Python 3.13 with NumPy 2.5, SciPy 1.18, pandas 3.0 and pyarrow 26 (the versions are recorded in the log of the confirmatory run). Every number in Sects.~\ref{sec:carrier} and~\ref{sec:fly} is printed in a result file of the repository, and each script names the file it writes. A validation report (Supplementary material) lists, for every table, the script and result file it comes from, the reruns of the scripts from the archived release with the hashes of their outputs, and the checks of every reference against its publisher record.

\paragraph{Use of a large language model.} The text of this manuscript was drafted by Claude, a large language model of Anthropic, from the author's earlier version of the work and under his direction, and was revised by the author, who decided the hypothesis, its wording, the scope of the paper and every analysis to be run. Claude also wrote and ran the analysis code, including the frozen code of the registration, which was run once after the registration was public; read the data points from the published figure; and searched for references and checked their bibliographic records. The author approved the registered thresholds and the reading of the BANC annotation table by names only before the registration was frozen, deposited the registration and every release of the repository himself, and reviewed and approved the full text. The checks of every number against its result file, the reruns of the scripts from a clean copy of the archived release, an independent recomputation of the registered statistics, and the checks of every reference against its publisher record were done in a separate pass and are reported in the validation report (Supplementary material). The model was Claude Opus (Claude Opus 5.5 for this version), accessed through claude.ai and Claude Code in November 2026.

\section*{Declarations}

\paragraph{Funding.} No funding was received for conducting this study.

\paragraph{Competing interests.} The author has no relevant financial or non-financial interests to disclose.

\paragraph{Ethics approval and consent.} Not applicable. The study uses published connectome releases and published measurements only; no experiments on humans or animals were done.

\paragraph{Data availability.} The registration, with the frozen code, the type rules, the hashes of the input files, the criteria and the decision rule, is archived at \url{https://doi.org/10.5281/zenodo.24223802}. The confirmatory result file, the run log and the deviations file are archived at \url{https://doi.org/10.5281/zenodo.24224168}. The exploratory and post hoc scripts, their result files and the points read from Fig.~4 of \citet{Seelig2015} are archived at \url{https://doi.org/10.5281/zenodo.24241203} (release v0.3.0); the analyses added for this version are in release v0.4.0 of the same record, which the concept DOI \url{https://doi.org/10.5281/zenodo.24223801} resolves to (all versions). The connectomes are public releases: male CNS v1.0, FlyWire v783 and BANC release 888. No new experimental data were generated.

\paragraph{Code availability.} All code is in the repository \url{https://github.com/erankopel/compartment-screen} and in the archived releases above.

\paragraph{Author contributions.} E.K. conceived the study, designed the analyses and the registration, directed the drafting, and reviewed, revised and approved the manuscript.

\paragraph{Preprint.} An earlier version of this work is available as arXiv:2611.23320 (\url{https://doi.org/10.48550/arXiv.2611.23320}), under a CC BY 4.0 licence.
\appendix
\section{Proof of the record bound}\label{app:proof}

The proof is that of \citet{Kopel2026b}, given here in full so that the record condition does not rest on a preprint.

\paragraph{Setting.} One round acts on a qubit label as a channel $\Phi$ with Stinespring isometry $V:\mathcal H_M\to\mathcal H_M\otimes\mathcal H_E$, so that $\Phi(\rho)=\mathrm{Tr}_E\,V\rho V^\dagger$. The registers $E$ are those that the round discards; what they receive is the complementary channel $\Phi^c(\rho)=\mathrm{Tr}_M\,V\rho V^\dagger$. Let $|\Omega\rangle = (|00\rangle+|11\rangle)/\sqrt2$, $J(\Phi)=(\mathrm{id}\otimes\Phi)(|\Omega\rangle\langle\Omega|)$ the normalised Choi state and $J^\Gamma$ its partial transpose on the reference. A qubit channel is entanglement-breaking if and only if $J(\Phi)$ is separable \citep{Horodecki2003}, which for two qubits holds if and only if $J^\Gamma\ge0$ \citep{Peres1996,Horodecki1996}. We use the signed margin $\mu(\Phi) = -\lambda_{\min}(J^\Gamma)$, so that $\Phi$ fails to be entanglement-breaking exactly when $\mu>0$. Fidelity is $F(\rho,\sigma)=\|\sqrt\rho\sqrt\sigma\|_1$ and trace distance $T(\rho,\sigma)=\frac12\|\rho-\sigma\|_1$.

\begin{lemma}\label{lem:uhlmann}
Let $|e_0\rangle,|e_1\rangle$ be orthonormal. Then
\begin{equation*}
\|\Phi(|e_1\rangle\langle e_0|)\|_1 = F\bigl(\Phi^c(|e_0\rangle\langle e_0|),\Phi^c(|e_1\rangle\langle e_1|)\bigr).
\end{equation*}
\end{lemma}
\begin{proof}
The vectors $|v_j\rangle=V|e_j\rangle$ purify $\rho_E^j=\Phi^c(|e_j\rangle\langle e_j|)$, with the label register as purifying system, and $\Phi(|e_1\rangle\langle e_0|)=\mathrm{Tr}_E|v_1\rangle\langle v_0|$. For two purifications on a common space, Uhlmann's theorem in the form $F(\rho_E^0,\rho_E^1)=\|\mathrm{Tr}_E|v_1\rangle\langle v_0|\|_1$ \citep{Uhlmann1976,Watrous2018} gives the claim.
\end{proof}

\begin{theorem}\label{thm:record}
Let $\{|e_0\rangle,|e_1\rangle\}$ be any orthonormal basis and $\rho_E^j=\Phi^c(|e_j\rangle\langle e_j|)$. Then
\begin{equation}
\mu(\Phi)\le\tfrac12\|\Phi(|e_1\rangle\langle e_0|)\|_\infty\le\tfrac12F(\rho_E^0,\rho_E^1).
\end{equation}
\end{theorem}
\begin{proof}
Let $W$ be the unitary with $W|j\rangle=|e_j\rangle$ and $\Phi_W=\Phi\circ\mathrm{Ad}_W$. Since $(1\otimes W)|\Omega\rangle=(W^{T}\otimes1)|\Omega\rangle$, $J(\Phi_W)^\Gamma=(W^\dagger\otimes1)J(\Phi)^\Gamma(W\otimes1)$, so $\mu(\Phi_W)=\mu(\Phi)$; moreover $\Phi_W(|1\rangle\langle0|)=\Phi(|e_1\rangle\langle e_0|)$ and $\Phi_W^c(|j\rangle\langle j|)=\rho_E^j$. It therefore suffices to take $|e_j\rangle=|j\rangle$. In the computational basis of the reference,
\begin{equation}
J^\Gamma=\tfrac12\sum_{m,n}|n\rangle\langle m|\otimes\Phi(|m\rangle\langle n|)=\tfrac12(D+O),
\end{equation}
with $D=\sum_m|m\rangle\langle m|\otimes\Phi(|m\rangle\langle m|)\ge0$ and $O=|0\rangle\langle1|\otimes\Phi(|1\rangle\langle0|)+|1\rangle\langle0|\otimes\Phi(|0\rangle\langle1|)$. The Hermitian matrix $O$ has eigenvalues $\pm s_k$, where $s_k$ are the singular values of $\Phi(|1\rangle\langle0|)$. By Weyl's inequality, $\lambda_{\min}(J^\Gamma)\ge\frac12[\lambda_{\min}(D)+\lambda_{\min}(O)]\ge-\frac12\|\Phi(|1\rangle\langle0|)\|_\infty$. The second inequality follows from $\|\cdot\|_\infty\le\|\cdot\|_1$ and Lemma~\ref{lem:uhlmann}.
\end{proof}

\begin{corollary}\label{cor:record}
Let $\mu_+=\max\{\mu,0\}$ and $T_E=T(\rho_E^0,\rho_E^1)$. Then (i) $(2\mu_+)^2+T_E^2\le1$; (ii) the bounds remain valid when $\rho_E^j$ are replaced by their reductions to any subset of the discarded registers; (iii) if the discarded registers, or any subset of them, distinguish $|e_0\rangle$ from $|e_1\rangle$ perfectly, $\Phi$ is entanglement-breaking.
\end{corollary}
\begin{proof}
(i) follows from $F\le(1-T^2)^{1/2}$ \citep{Fuchs1999}. (ii) Fidelity does not decrease under partial traces. (iii) If $F=0$, Lemma~\ref{lem:uhlmann} gives $\Phi(|e_1\rangle\langle e_0|)=0$, so $\Phi=\Phi\circ\Delta_e$ with $\Delta_e$ the complete dephasing in the basis $\{|e_j\rangle\}$. This is a measure-and-prepare channel, hence entanglement-breaking \citep{Horodecki2003}.
\end{proof}

\begin{remark}
The bounds are tight. For pure dephasing, $V|j\rangle=|j\rangle\otimes|\chi_j\rangle$, one finds $\mu=\frac12|\langle\chi_0|\chi_1\rangle|=\frac12F(\rho_E^0,\rho_E^1)$, and since the register states are pure, $(2\mu)^2+T_E^2=1$. For a $d$-level label the same argument, with Schur's test on the off-diagonal blocks of $J^\Gamma$, gives $\lambda_{\min}(J^\Gamma)\ge-\frac1d\max_n\sum_{m\ne n}F(\rho_E^m,\rho_E^n)$, and part (iii) holds when the discarded registers distinguish all $d$ basis states.
\end{remark}

\paragraph{Rounds.} A memoryless loop with fresh registers in every round acts as $\Phi^n$ after $n$ rounds, and the registers discarded by $\Phi^n$ are those of all $n$ rounds. Their reduction to the first $n-1$ rounds is the complementary output of $\Phi^{n-1}$, so by Corollary~\ref{cor:record}(ii) the fidelity $F_n$ of the $n$-round record is non-increasing in $n$ and the bound $\mu(\Phi^n)\le\frac12F_n$ can only tighten as rounds accumulate. How fast it tightens depends on the loop. The round counts quoted in Sect.~\ref{sec:record} (at most 3 rounds with $T\ge0.99$ in every round, 11 with $T\ge0.9$; also 5 with $T\ge0.95$ and 2 with $T\ge0.999$) are maxima over 20,000 sampled loops of the circuit family of \citet{Kopel2026b} and are not certified.

\section{Walk measures}\label{app:walk}

\paragraph{Graph and walk.} The graph holds every neuron of the data set (in BANC and FlyWire, every row of the annotation table whose super class is not glia, trachea or not a neuron; in the male, the neurons of the released neuron-to-neuron edge table) and every connection between two of them with at least a threshold number of synapses $\theta$ (primary analysis $\theta=1$), weighted by its synapse count $w_{ij}$. The walk runs on the giant strongly connected component, with $P_{ij}=w_{ij}/\sum_k w_{ik}$ (rows presynaptic) and stationary distribution $\pi$ found by power iteration.

\paragraph{Stopped and conditioned walk.} For a set $S$, $Q=P_{SS}$ is substochastic. The share lost per step is $\sum_{i\in S}\pi_i(1-\sum_{j\in S}P_{ij})/\sum_{i\in S}\pi_i$. With $\lambda_1$ the real part of the eigenvalue of $Q$ of largest modulus and $\lambda_2$ the modulus of the next, $t_{\text{stay}}=1/(1-\lambda_1)$ and $t_{\text{relax}}=1/(1-|\lambda_2|/\lambda_1)$. Let $h$ be the right Perron vector of $Q$; neurons with $h_i=0$, from which a walker cannot remain in $S$, are dropped and counted. The walk conditioned to stay in $S$ for ever, the Doob transform of $Q$, has $\hat Q_{ij}=Q_{ij}h_j/(\lambda_1h_i)$; it is stochastic, and its stationary distribution is $\nu_i\propto\ell_ih_i$, with $\ell$ the left Perron vector. The Dobrushin coefficient $\delta_r$ is computed over all pairs of starting neurons.

\paragraph{Finite-time mixing.} $d(t)=\max_x\|\hat Q^{\,t}(x,\cdot)-\nu\|_{\text{TV}}$, over all starting neurons for sets up to 1,600 neurons and over 400 seeded starts otherwise, and $t_{1/4}$ is the first $t$ with $d(t)\le\frac14$ \citep{Levin2017}. Because $\hat Q$ is not reversible, its relaxation time need not track $d(t)$: for a matrix far from normal, $\|\hat Q^{\,t}-\mathbf 1\nu\|$ can stay large for many steps even when every eigenvalue but the leading one is small \citep{Trefethen2005}. Henrici's departure from normality, scaled to $[0,1]$, is $\bigl(\|\hat Q\|_F^2-\sum_i|\lambda_i(\hat Q)|^2\bigr)^{1/2}/\|\hat Q\|_F$ \citep{Henrici1962}.

\paragraph{Exit channel.} For two starting neurons $x,y\in S$ the exit record is the pair (first step at which the walker is outside $S$, the neuron it lands on). Its law is computed exactly by propagating the walk on $S$ and collecting the flow onto outside neurons at every step, until less than $10^{-4}$ of the mass remains. The exit distance of $x$ and $y$ is the total variation distance between their two laws; 240 seeded starts give 28,680 pairs per set. By the strong Markov property the future of the walk after the exit depends on the start only through the exit record, so no observer of the walk outside $S$ can distinguish $x$ from $y$ better than the exit distance allows.

\paragraph{Excitatory walk.} Only connections whose presynaptic neuron is predicted to be cholinergic \citep{Eckstein2024} are kept, and the walk is rebuilt on the giant strongly connected component of that graph. Neurons without cholinergic output, the GABAergic ring neurons among them, leave the component. Predicted transmitters are the consensus prediction per neuron in the male and the per-neuron prediction of the annotation table in FlyWire and BANC; missing or unclear predictions count as unknown.

\paragraph{Signed linear rate model.} On the full walk, $x(t+1)=x(t)A$ with $A=\mathrm{diag}(s)P$. Scheme S1: $s_i=+1$ for acetylcholine, $-1$ for GABA, glutamate and histamine, $0$ for amines and unknown. Scheme S2: as S1 with glutamate and unknown set to $+1$. For each set $S$, with $\rho$ the largest eigenvalue modulus of $A_{SS}$ and $\mu_2$ the largest modulus other than the leading eigenvalue and its complex conjugate, persistence is $1/(1-\rho)$ and internal relaxation $1/(1-\mu_2/\rho)$. The model has no nonlinearity, no time constants beyond the step, and no neuromodulation.

\paragraph{In silico silencing.} A silenced neuron passes nothing on: a walker that steps onto it is lost. The EB set with a family $X$ silenced is therefore the set $\mathrm{EB}\setminus X$ under the unchanged $P$, and its statistics are computed as above. As a control the same number of EB neurons is removed at random, ten times.

\section{Points read from the published figure}\label{app:points}

\paragraph{Reading.} Figure~4 of \citet{Seelig2015} is stored as vector graphics in the published file. The markers of panels h and i were read with their plotted coordinates and calibrated from the tick marks. Each panel holds 445 markers; the legend prints $n=499$. The 445 points give a duration of $6.71\pm5.14$~s and a change of $0.020\pm0.762$~rad, and panel h gives $r=0.703$, which are the printed statistics. The markers of the two panels come in the same order, and the difference computed from panel h equals the value plotted in panel i for every bout to within 0.004~rad.

\paragraph{Two estimates.} The pooled bound treats the whole spread as diffusion over bouts of random length, which gives $D\le0.043$~rad$^2$/s. The mixture model was fitted by maximum likelihood to the single bouts: with probability $p$ the position after the bout is unrelated to the one before, and otherwise the change is normal with variance $s_0^2+2Dt$. The fit gives $p=0.105$ (95\% bootstrap interval 0.060 to 0.154), $s_0=0.32$~rad (0.10 to 0.44) and $D=0.012$~rad$^2$/s (0.0049 to 0.020). Without diffusion the fit is worse (likelihood ratio $p=6\times10^{-5}$); without the unrelated share it is far worse; and an unrelated share that grows with duration is no better ($p=0.78$). The model was chosen after looking at the points. The three further panels in darkness of the same paper (Extended Data Fig.~10b, d, f) give pooled values of 0.052 to 0.076~rad$^2$/s.

\paragraph{Mutual information.} With the initial heading uniform on the circle, the information that the position after $t$ seconds keeps about it is $I(t)=\log_2(2\pi)-h(f_t)$, where $f_t=(1-p)\,\mathrm{WN}(s_0^2+2Dt)+p/(2\pi)$ is the density of the change on the circle, $\mathrm{WN}(\sigma^2)$ the wrapped normal density with variance $\sigma^2$, and $h$ the differential entropy in bits \citep{Cover2006}. The integral is evaluated on a grid of 40,000 points. The crossing times of Table~\ref{tab:info} are found by bisection. For the pooled bound ($s_0=0$, $p=0$) $I$ is unbounded at $t\to0$; at the count-based hold time of 7.2~s for four positions it is 0.95 bits.

\paragraph{Limits.} The points were read from a figure. The unrelated share may be jumps, fading and return of the bump, or turning during the 0.59~s after the restart over which the second estimate is averaged. All estimates are for flies standing on a ball in darkness between walking bouts of 2.6 to 36~s. In tethered flight in darkness the bump drifts within seconds \citep{Kim2017}.

\clearpage
\section*{Supplementary material}
\addcontentsline{toc}{section}{Supplementary material}
\setcounter{table}{0}\renewcommand{\thetable}{S\arabic{table}}
\noindent Sections S1 to S11 and Tables S1 to S17. Section, table and appendix numbers without S refer to this article. Sections S1 to S9 are carried over from the earlier version of this work (arXiv:2611.23320v1), with the terms of the present version; Sections S10 and S11 are new. Everything after the registered run of 14 November 2026 is exploratory. The works cited are listed in the references of this article.

\hypertarget{s1.-what-is-taken-from-the-source-theory}{%
\section*{S1. What is taken from the source
theory}}

\begin{longtable}[]{@{}
  >{\raggedright\arraybackslash}p{(\columnwidth - 4\tabcolsep) * \real{0.4000}}
  >{\raggedright\arraybackslash}p{(\columnwidth - 4\tabcolsep) * \real{0.3000}}
  >{\raggedright\arraybackslash}p{(\columnwidth - 4\tabcolsep) * \real{0.3000}}@{}}
\toprule\noalign{}
\begin{minipage}[b]{\linewidth}\raggedright
Element in Lahav and Neemeh (2022)
\end{minipage} & \begin{minipage}[b]{\linewidth}\raggedright
Status here
\end{minipage} & \begin{minipage}[b]{\linewidth}\raggedright
Reason
\end{minipage} \\
\midrule\noalign{}
\endhead
\bottomrule\noalign{}
\endlastfoot
Three-part system S = (T, P, C), Eqs. 2 to 10 & Kept & Gives the loop a
definite shape \\
C takes its own output as input, Eqs. 22 and 42 & Kept, central & This
is the loop under test \\
Outside access only through the sensing subsystem, Eqs. 36 to 43 & Kept
& Basis of the boundary condition \\
Position condition in the transformation, Eq. 46 & Kept in weak form &
Readout must be local to the system \\
Judgment equals quale in the system's own frame, Eq. 35 & Not used;
experience is not claimed & It is argued, not measured \\
Analogy with the Lorentz transformation, Eqs. 1 and 44 & Dropped & Not
needed for any condition \\
Claim that the hard problem dissolves & Dropped & Outside the scope of
necessary conditions \\
\end{longtable}

\noindent\textbf{Table S1.} Elements of the source theory and their status in this paper.

\hypertarget{s2.-the-floor-in-numbers}{%
\section*{S2. The floor in numbers}}

\begin{longtable}[]{@{}
  >{\raggedright\arraybackslash}p{(\columnwidth - 2\tabcolsep) * \real{0.4667}}
  >{\raggedright\arraybackslash}p{(\columnwidth - 2\tabcolsep) * \real{0.5333}}@{}}
\toprule\noalign{}
\begin{minipage}[b]{\linewidth}\raggedright
Quantity
\end{minipage} & \begin{minipage}[b]{\linewidth}\raggedright
Value at 310 K
\end{minipage} \\
\midrule\noalign{}
\endhead
\bottomrule\noalign{}
\endlastfoot
kT & 4.28 $\times$ 10\textsuperscript{$-$21} J, or 26.7 meV \\
Energy floor, kT ln 2 & 2.97 $\times$ 10\textsuperscript{$-$21} J per bit \\
Speed floor, h / 4E at E = kT & 3.9 $\times$ 10\textsuperscript{$-$14} s \\
The 200 fs limit of Fields et al.~(2021) & 5.2 speed floors \\
Bits a 20 W brain can rewrite & 6.7 $\times$ 10\textsuperscript{21} per second,
or 7.8 $\times$ 10\textsuperscript{10} per neuron per second \\
Allowance for one 25 ms round & 1.7 $\times$ 10\textsuperscript{20} bits for
the brain, 2 $\times$ 10\textsuperscript{9} per neuron \\
Allowance for one 200 fs round & 1.3 $\times$ 10\textsuperscript{9} bits for
the whole brain, 0.016 bits per neuron \\
Planck capacity ceiling, brain-sized surface & 8 $\times$
10\textsuperscript{67} bits \\
\end{longtable}

\noindent\textbf{Table S2.} The floor in numbers. The biological inputs are textbook
orders of magnitude: 20 W, 8.6 $\times$ 10\textsuperscript{10} neurons,
10\textsuperscript{14} to 10\textsuperscript{15} synapses.

A screen of one bit per synapse can be rewritten at most 7 $\times$
10\textsuperscript{6} to 7 $\times$ 10\textsuperscript{7} times per second, and
one of one bit per neuron at most 8 $\times$ 10\textsuperscript{10} times per
second. The floor is not set at the Planck scale: the Planck time lies
some 42 orders of magnitude below a neural round of 0.1 s, and the
Planck capacity ceiling is about 53 orders of magnitude above the
synapse count, so neither decides anything.

\hypertarget{s3.-record-condition-check-of-the-bound-and-rounds-against-record}{%
\section*{S3. Record condition: check of the bound, and rounds against
record}}

The loop of Kopel (2026b) was rebuilt from its circuit, without the
closed forms. The Bloch data agree with the closed forms to
10\textsuperscript{$-$15} on 2,000 random loops, and at the reference
point the circuit gives the published values, including an index of 3.
On 20,000 random loops, with the fidelity F and the trace distance T
computed from the states of the discarded registers, there is no
violation of $\mu$ $\le$ F/2 or of (2$\mu$)\textsuperscript{2} + T\textsuperscript{2} $\le$ 1. Both bounds are approached to
within 10\textsuperscript{$-$8}.

\begin{longtable}[]{@{}
  >{\raggedright\arraybackslash}p{(\columnwidth - 6\tabcolsep) * \real{0.2101}}
  >{\raggedright\arraybackslash}p{(\columnwidth - 6\tabcolsep) * \real{0.2017}}
  >{\raggedright\arraybackslash}p{(\columnwidth - 6\tabcolsep) * \real{0.4118}}
  >{\raggedright\arraybackslash}p{(\columnwidth - 6\tabcolsep) * \real{0.1765}}@{}}
\toprule\noalign{}
\begin{minipage}[b]{\linewidth}\raggedright
Record per round, T at least
\end{minipage} & \begin{minipage}[b]{\linewidth}\raggedright
Bound on $\mu$ after one round
\end{minipage} & \begin{minipage}[b]{\linewidth}\raggedright
Loops entanglement-breaking at round 1
\end{minipage} & \begin{minipage}[b]{\linewidth}\raggedright
Largest index found
\end{minipage} \\
\midrule\noalign{}
\endhead
\bottomrule\noalign{}
\endlastfoot
0.2 & 0.490 & 24\% & 109 \\
0.6 & 0.400 & 28\% & 21 \\
0.8 & 0.300 & 34\% & 11 \\
0.9 & 0.218 & 42\% & 11 \\
0.95 & 0.156 & 51\% & 5 \\
0.99 & 0.071 & 65\% & 3 \\
0.999 & 0.022 & 80\% & 2 \\
1 & 0 & 100\% & 1 \\
\end{longtable}

\noindent\textbf{Table S3.} Largest entanglement-breaking index found among 20,000 random
loops when the discarded registers hold a record with trace distance at
least T in every round. These are sample maxima for one loop family, not
certified bounds.

\hypertarget{s4.-boundary-condition-the-boundary-bit-with-a-frame-mismatch}{%
\section*{S4. Boundary condition: the boundary bit with a frame
mismatch}}

One side prepares a bit in its own frame and the other measures it in a
frame rotated by an angle $\theta$. The result is a binary symmetric channel
with flip probability sin\textsuperscript{2}($\theta$/2). Its Dobrushin coefficient is
\textbar cos $\theta$\textbar, which held for each of 20,000 sampled pairs of
inputs, and its contraction coefficient for relative entropy is cos\textsuperscript{2} $\theta$.
After r rounds the two are \textbar cos $\theta$\textbar{}\textsuperscript{r}
and cos\textsuperscript{2r} $\theta$, checked by matrix powers up to r = 64.

\begin{longtable}[]{@{}
  >{\raggedright\arraybackslash}p{(\columnwidth - 6\tabcolsep) * \real{0.2500}}
  >{\raggedright\arraybackslash}p{(\columnwidth - 6\tabcolsep) * \real{0.2583}}
  >{\raggedright\arraybackslash}p{(\columnwidth - 6\tabcolsep) * \real{0.2417}}
  >{\raggedright\arraybackslash}p{(\columnwidth - 6\tabcolsep) * \real{0.2500}}@{}}
\toprule\noalign{}
\begin{minipage}[b]{\linewidth}\raggedright
Mismatch $\theta$
\end{minipage} & \begin{minipage}[b]{\linewidth}\raggedright
Kept per round, cos\textsuperscript{2} $\theta$
\end{minipage} & \begin{minipage}[b]{\linewidth}\raggedright
Rounds to lose half
\end{minipage} & \begin{minipage}[b]{\linewidth}\raggedright
Rounds to fall to 1\%
\end{minipage} \\
\midrule\noalign{}
\endhead
\bottomrule\noalign{}
\endlastfoot
1$^{\circ}$ & 0.9997 & 2,275 & 15,117 \\
5$^{\circ}$ & 0.992 & 91 & 604 \\
10$^{\circ}$ & 0.970 & 23 & 150 \\
20$^{\circ}$ & 0.883 & 5.6 & 37 \\
30$^{\circ}$ & 0.750 & 2.4 & 16 \\
45$^{\circ}$ & 0.500 & 1.0 & 6.6 \\
60$^{\circ}$ & 0.250 & 0.5 & 3.3 \\
\end{longtable}

\noindent\textbf{Table S4.} Survival of a boundary bit under a frame mismatch.

After r rounds the outside holds (cos $\theta$\textsubscript{out} / cos
$\theta$\textsubscript{in})\textsuperscript{2r} of what the inside holds, which
is 0.13 after 8 rounds for 10$^{\circ}$ inside and 30$^{\circ}$ outside. For a screen of N
independent bits the relative-entropy coefficient stays at cos\textsuperscript{2} $\theta$, while
the Dobrushin coefficient climbs towards 1: at 60$^{\circ}$ it is 0.50, 0.69,
0.86, 0.97 and 1.00 for N = 1, 4, 8, 16 and 64. The boundary condition
on a wide screen should therefore be stated with the relative-entropy
coefficient. The Dobrushin coefficient stays the right tool for the
single-walker chain of the compartment condition.

\hypertarget{s5.-data-graph-and-sets}{%
\section*{S5. Data, graph and sets}}

\textbf{Data.} Male CNS v1.0 (Berg et al., 2026), FlyWire v783
(Dorkenwald et al., 2024; Schlegel et al., 2024) and BANC release 888
(Bates et al., 2026). For BANC the compiled annotation table and the
simple edge list (v3) were taken from the public bucket and fixed by
SHA-256 in the registration.

\textbf{Graph.} Every row of the annotation table with a super class
other than glia, trachea or not\_a\_neuron, and every connection with
both ends among them, weighted by its synapse count. The walk runs on
the giant strongly connected component. The primary analysis uses all
connections.

\textbf{Sets, by name only.} EB is every neuron whose type matches a
family rule: ring neurons ER1 to ER6, EPG, PEG, PEN, EL, ExR1, 2, 3, 4
and 6, TuBu, MeTu2, LPsP, GLNO, AOTU046 and TuTuB. CX is the three
central-complex classes plus EB. MB is Kenyon cells with mushroom-body
output and dopaminergic neurons. AL is antennal-lobe projection and
local neurons. The EB rule selects 691 neurons in the male, 683 in
FlyWire and 595 in BANC.

\textbf{Names-only check of BANC before the freeze.} Read from the
annotation table: the column names, the name columns, and counts of
region, status and proofreading. No connectivity. The table has 188,508
rows, of which 146,769 form the graph. All class names resolve (CX
3,164, MB 4,957, AL 1,128) and every family rule finds a match. One rule
changed as a result: the super class not\_a\_neuron (195 rows) is
excluded with glia and trachea. One limitation was registered: the
visual input types are less completely typed in BANC (MeTu2 has 27
neurons against 101 in the male and in FlyWire), and the rule is applied
as it stands.

\textbf{Statistics.} Lost per step is the share of the stationary mass
of a set that leaves it in one step. Stay and mixing times are defined
in Sect. 5 and Appendix B of the main text. The Dobrushin coefficient
inside a set is computed for the walk conditioned to remain in it. The
exit channel uses 240 seeded starting neurons per set and the total
variation distance between their exit records (exit step and exit
neuron). The reach ratio from A to B is the probability that a walker
leaving A is in B after one step, over the same probability for a
stationary walker.

\hypertarget{s6.-exploratory-results-on-two-animals}{%
\section*{S6. Exploratory results on two
animals}}

All results of this section are exploratory.

\begin{longtable}[]{@{}
  >{\raggedright\arraybackslash}p{(\columnwidth - 14\tabcolsep) * \real{0.1833}}
  >{\raggedright\arraybackslash}p{(\columnwidth - 14\tabcolsep) * \real{0.1083}}
  >{\raggedright\arraybackslash}p{(\columnwidth - 14\tabcolsep) * \real{0.1000}}
  >{\raggedright\arraybackslash}p{(\columnwidth - 14\tabcolsep) * \real{0.1083}}
  >{\raggedright\arraybackslash}p{(\columnwidth - 14\tabcolsep) * \real{0.1250}}
  >{\raggedright\arraybackslash}p{(\columnwidth - 14\tabcolsep) * \real{0.1000}}
  >{\raggedright\arraybackslash}p{(\columnwidth - 14\tabcolsep) * \real{0.1500}}
  >{\raggedright\arraybackslash}p{(\columnwidth - 14\tabcolsep) * \real{0.1250}}@{}}
\toprule\noalign{}
\begin{minipage}[b]{\linewidth}\raggedright
Set (male)
\end{minipage} & \begin{minipage}[b]{\linewidth}\raggedright
Neurons
\end{minipage} & \begin{minipage}[b]{\linewidth}\raggedright
Lost per step
\end{minipage} & \begin{minipage}[b]{\linewidth}\raggedright
Stays (steps)
\end{minipage} & \begin{minipage}[b]{\linewidth}\raggedright
Mixes inside (steps)
\end{minipage} & \begin{minipage}[b]{\linewidth}\raggedright
Stays / mixes
\end{minipage} & \begin{minipage}[b]{\linewidth}\raggedright
Exit distance: largest, mean
\end{minipage} & \begin{minipage}[b]{\linewidth}\raggedright
Exit pairs below 0.5
\end{minipage} \\
\midrule\noalign{}
\endhead
\bottomrule\noalign{}
\endlastfoot
EB set (sweep cut), all connections & 727 & 0.116 & 9.3 & 4.8 & 1.9 &
0.78, 0.30 & 89\% \\
Same set, connections of 5 synapses or more & 727 & 0.102 & 10.2 & 6.0 &
1.7 & & \\
Whole central-complex class & 2,950 & 0.113 & 12.3 & 9.4 & 1.3 & 0.97,
0.53 & 46\% \\
Central complex outside the EB set & 2,524 & 0.151 & 5.8 & 6.7 & 0.9 &
0.98, 0.52 & 54\% \\
Mushroom body & 4,501 & 0.367 & 3.5 & 16.5 & 0.2 & 0.99, 0.50 & 45\% \\
Kenyon cells alone & 4,064 & 0.486 & 2.3 & 56 & 0.04 & 0.99, 0.71 &
19\% \\
Antennal lobe & 1,106 & 0.473 & 2.8 & 13.6 & 0.2 & 1.00, 0.83 & 4\% \\
Random sets of 727 neurons, 20 draws & 727 & 0.995 to 0.998 & about 1 &
& & & \\
\end{longtable}

\noindent\textbf{Table S5.} Compartment statistics and exit channel for all sets tested in
the male.

\begin{itemize}
\item
  \textbf{A recurrent core with a feedforward input layer.} The 277
  neurons of the visual pathway carry 10\% of the inflow and hold a
  walker for under 2 steps. The 450 core neurons alone give nearly the
  same numbers as the whole set (stays 9.1 steps, mixes in 4.9).
\item
  \textbf{One interface.} 90\% of the inflow lands on 214 neurons, and
  90\% of the outflow leaves from 109 of those same neurons, with the
  EPG compass neurons carrying 55\% of it. The main partner is the rest
  of the central complex (Delta7 takes 30\% of the outflow), not sensory
  input.
\item
  \textbf{Forgetting inside.} The Dobrushin coefficient of the
  conditioned walk over all pairs is 1.00, 0.77, 0.25, 0.04 and 0.001 at
  2, 4, 8, 16 and 32 steps.
\item
  \textbf{Exit channel in full.} Over all 263,901 pairs of the EB set
  the exit distance is at most 0.777, with mean 0.288 and median 0.268.
  Stopping at 96 steps adds at most 3 $\times$ 10\textsuperscript{$-$5}. Without
  the exit time the mean falls to 0.216. At any single step the part of
  the difference visible outside averages at most 0.13, while the part
  visible inside starts at 0.80 and is down to 0.06 by step 8.
\item
  \textbf{Null models.} Keeping the synapse counts between every pair of
  cell types and shuffling which neurons carry them reproduces every
  value of the EB set (stays / mixes 1.95 to 1.96, Dobrushin coefficient
  0.25 to 0.26, exit mean 0.268, coupling contrast 9.2). Keeping only
  the counts between superclasses erases them (lost per step 0.99,
  Dobrushin coefficient 1.00), and so does the configuration model.
\item
  \textbf{Mushroom body and central complex.} The union loses 0.169 of
  its mass per step against 0.113 for the central complex alone in the
  male, and 0.149 against 0.093 in FlyWire. A walker leaving the
  mushroom body reaches the central complex within 1 and 4 steps with
  probability 3.5\% and 11.5\%, against 0.28\% and 1.0\% for a
  stationary walker. The reverse direction gives 3.6\% and 9.5\% against
  0.29\% and 0.89\%. The antennal lobe starts at chance (0.4\% at one
  step).
\end{itemize}

\begin{longtable}[]{@{}
  >{\raggedright\arraybackslash}p{(\columnwidth - 10\tabcolsep) * \real{0.1833}}
  >{\raggedright\arraybackslash}p{(\columnwidth - 10\tabcolsep) * \real{0.2500}}
  >{\raggedright\arraybackslash}p{(\columnwidth - 10\tabcolsep) * \real{0.1000}}
  >{\raggedright\arraybackslash}p{(\columnwidth - 10\tabcolsep) * \real{0.1833}}
  >{\raggedright\arraybackslash}p{(\columnwidth - 10\tabcolsep) * \real{0.1333}}
  >{\raggedright\arraybackslash}p{(\columnwidth - 10\tabcolsep) * \real{0.1500}}@{}}
\toprule\noalign{}
\begin{minipage}[b]{\linewidth}\raggedright
Flow at the EB set (male)
\end{minipage} & \begin{minipage}[b]{\linewidth}\raggedright
Acetylcholine
\end{minipage} & \begin{minipage}[b]{\linewidth}\raggedright
GABA
\end{minipage} & \begin{minipage}[b]{\linewidth}\raggedright
Glutamate
\end{minipage} & \begin{minipage}[b]{\linewidth}\raggedright
Amines
\end{minipage} & \begin{minipage}[b]{\linewidth}\raggedright
Entropy (bits)
\end{minipage} \\
\midrule\noalign{}
\endhead
\bottomrule\noalign{}
\endlastfoot
Whole CNS & 41\% & 26\% & 19\% & 1.2\% & 1.96 \\
Into the set & 45\% & 9\% & 31\% & 7.5\% & 2.00 \\
Inside the set & 37\% & 45\% & 6\% & 11.4\% & 1.84 \\
Out of the set & 74\% & 5\% & 8\% & 12.6\% & 1.42 \\
\end{longtable}

\noindent\textbf{Table S6.} Transmitter composition of the flow, with the chemical
identity of a signal taken as the predicted transmitter of the
presynaptic neuron.

Chemistry gives at most about 2 bits per signal at this boundary, so the
identity of a signal has to come from wiring. The figure for the inside
leans on the prediction that the 18 EL neurons are octopaminergic, and
two serotonergic ExR3 neurons carry 8\% of the outflow. The amine
statistic does not carry across data sets: the transmitter predictions
of FlyWire mark 8,410 neurons as aminergic, Kenyon cells included,
against 541 in the male consensus. No amine criterion was registered for
that reason.

\begin{longtable}[]{@{}
  >{\raggedright\arraybackslash}p{(\columnwidth - 12\tabcolsep) * \real{0.1750}}
  >{\raggedright\arraybackslash}p{(\columnwidth - 12\tabcolsep) * \real{0.0833}}
  >{\raggedright\arraybackslash}p{(\columnwidth - 12\tabcolsep) * \real{0.1250}}
  >{\raggedright\arraybackslash}p{(\columnwidth - 12\tabcolsep) * \real{0.1250}}
  >{\raggedright\arraybackslash}p{(\columnwidth - 12\tabcolsep) * \real{0.1250}}
  >{\raggedright\arraybackslash}p{(\columnwidth - 12\tabcolsep) * \real{0.2250}}
  >{\raggedright\arraybackslash}p{(\columnwidth - 12\tabcolsep) * \real{0.1417}}@{}}
\toprule\noalign{}
\begin{minipage}[b]{\linewidth}\raggedright
Statistic
\end{minipage} & \begin{minipage}[b]{\linewidth}\raggedright
Male, all
\end{minipage} & \begin{minipage}[b]{\linewidth}\raggedright
Male, 5 synapses
\end{minipage} & \begin{minipage}[b]{\linewidth}\raggedright
FlyWire, all
\end{minipage} & \begin{minipage}[b]{\linewidth}\raggedright
FlyWire, 5 synapses
\end{minipage} & \begin{minipage}[b]{\linewidth}\raggedright
Where there is no compartment
\end{minipage} & \begin{minipage}[b]{\linewidth}\raggedright
Threshold
\end{minipage} \\
\midrule\noalign{}
\endhead
\bottomrule\noalign{}
\endlastfoot
EB lost per step & 0.116 & 0.102 & 0.094 & 0.074 & MB 0.37 to 0.66;
superclass null 0.99 & below 0.20 \\
EB stays / mixes & 1.92 & 1.70 & 2.80 & 2.49 & MB 0.21 to 0.64; AL 0.12
to 0.20; superclass null 0.03 to 0.08 & above 1.3 \\
EB Dobrushin coefficient, 8 steps & 0.255 & 0.364 & 0.210 & 0.381 &
Superclass null 1.00 & below 0.6 \\
EB exit channel, mean & 0.267 & 0.235 & 0.231 & 0.197 & MB 0.50 to 0.71;
superclass null 1.00 & below 0.37 \\
EB exit channel, pairs below 0.5 & 95\% & 98\% & 97\% & 99\% & MB 18\%
to 45\%; CX 33\% to 51\% & above 70\% \\
Reach MB to CX over AL to CX, one step & 9.2 & 10.8 & 13.4 & 17.2 &
Superclass null 1.2 & at least 3 \\
\end{longtable}

\noindent\textbf{Table S7.} Values behind the registered thresholds (name-based EB set).
Each threshold is the rounded geometric midpoint between the weakest of
the four exploratory values and the nearest value that marks no
compartment. The gap of 0.10 to MB in C4 is half the smallest gap seen.
All criteria hold in FlyWire, which is partly circular because FlyWire
helped to set them.

\hypertarget{s7.-registration-protocol-and-values-reported-but-not-scored}{%
\section*{S7. Registration, protocol and values reported but not
scored}}

\textbf{Timeline, 14 Nov 2026 (UTC).} Registration archived 15:59.
Registered analysis run once, 16:03 to 16:07. Results archived 16:22.

\textbf{Protocol.} One run of the frozen files (the library, the run
script and the type rules, each fixed by hash), no rerun, deviations
logged. The compartment hypothesis is confirmed only if C1 to C4 all
hold with all connections; C5 is judged on its own; every number is
reported whatever the outcome.

\textbf{Deviations.} None from the registered analysis. The deviations
file records four things the registration does not describe: the
registered command was started through a wrapper that writes time
stamps, library versions and file hashes to the log; the annotation
table was downloaded again for the run, with the same hash; the patch
version of the interpreter; and an exit-code line in the log that
reports the status of a preceding command, so that completion is shown
by the final log line and the complete result file.

\begin{longtable}[]{@{}
  >{\raggedright\arraybackslash}p{(\columnwidth - 4\tabcolsep) * \real{0.4000}}
  >{\raggedright\arraybackslash}p{(\columnwidth - 4\tabcolsep) * \real{0.3417}}
  >{\raggedright\arraybackslash}p{(\columnwidth - 4\tabcolsep) * \real{0.2583}}@{}}
\toprule\noalign{}
\begin{minipage}[b]{\linewidth}\raggedright
Statistic (BANC)
\end{minipage} & \begin{minipage}[b]{\linewidth}\raggedright
All connections
\end{minipage} & \begin{minipage}[b]{\linewidth}\raggedright
5 synapses or more
\end{minipage} \\
\midrule\noalign{}
\endhead
\bottomrule\noalign{}
\endlastfoot
Giant component, neurons & 144,122 & 113,971 \\
EB neurons in it & 589 & 555 \\
EB lost per step & 0.133 & 0.125 \\
EB stays / mixes & 1.396 & 0.84 \\
EB Dobrushin coefficient at 4, 8, 16 steps & 0.75, 0.41, 0.12 & 0.98,
0.90, 0.73 \\
EB exit channel: mean, pairs below 0.5, largest & 0.307, 87\%, 1.00 &
0.360, 78\%, 1.00 \\
MB lost per step; stays / mixes & 0.311; 0.35 & 0.709; 2.86 \\
AL lost per step; stays / mixes & 0.423; 0.19 & 0.277; 0.40 \\
MB exit channel, mean & 0.582 & 0.719 \\
CX stays / mixes & 0.79 & 0.36 \\
Union of CX and MB, lost per step & 0.190 & 0.228 \\
Reach MB to CX over AL to CX & 12.1 & 8.3 \\
Reach CX to MB over a stationary walker & 6.7 & 1.6 \\
\end{longtable}

\noindent\textbf{Table S8.} Registered statistics with all connections (scored) and with
connections of at least 5 synapses (reported, not scored). At 5 synapses
the mushroom-body set shrinks to 2,147 of its 4,898 neurons in the
component, which is why its ratio is no longer comparable.

Amine shares by type name, reported and not scored: 0.2\% of the flow in
the whole graph, 0.6\% of the flow into the EB set, 7.9\% inside it and
14.4\% out of it.

\hypertarget{s8.-post-hoc-analyses-of-the-banc-result}{%
\section*{S8. Post hoc analyses of the BANC
result}}

Everything in this section was done after the confirmatory run. It is
exploratory: nothing was registered, no criterion is scored, the
registered verdict is unchanged, and the confirmatory analysis was not
rerun.

\hypertarget{s8.1-where-the-criteria-first-fail}{%
\subsection*{S8.1 Where the criteria first
fail}}

In BANC stays / mixes falls below 1.3 at threshold 2 and the Dobrushin
coefficient passes 0.6 at threshold 3 (Table 5 of the main text). In the
male and in FlyWire the first failing thresholds are 15 for stays /
mixes, and 12 and 10 for the Dobrushin coefficient. Above threshold 10
the BANC values are not interpretable, because the set breaks up.

\hypertarget{s8.2-scale}{%
\subsection*{S8.2 Scale}}

Synapses per neuron in the edge lists: 268 in BANC, 745 in the male, 495
in FlyWire. Inside the EB set, per neuron: 434, 1,213 and 669. The
median connection inside the set has 3, 5 and 4 synapses.

\hypertarget{s8.3-density-or-pattern}{%
\subsection*{S8.3 Density or pattern}}

\begin{itemize}
\item
  \textbf{Uniform thinning.} With every synapse of the male kept at
  random with the probability that gives the BANC density, the male
  still has stays / mixes 1.51 and a Dobrushin coefficient of 0.59 at 5
  synapses. FlyWire thinned the same way keeps 3.3 and 0.42.
\item
  \textbf{Pattern thinning.} Against the male, family-to-family blocks
  keep between 5\% and 72\% of their synapses in BANC. Thinning block by
  block with that pattern gives stays / mixes 1.22, 0.88, 0.70 and 0.30
  in the male and 1.89, 1.65, 1.26 and 0.58 in FlyWire, at thresholds 1,
  2, 3 and 5.
\item
  \textbf{Which connections.} Ring neurons onto EL, ExR and EPG neurons
  keep 22\% of the male's synapses in BANC, and the return connections
  onto ring neurons 18\%, against 36\% overall. In both source animals,
  weakening the ring-neuron output alone raises the Dobrushin
  coefficient to or near the BANC values, and restoring it alone removes
  most of the effect. The visual input neurons, the least completely
  typed part of BANC, are not the cause.
\item
  \textbf{The slow mode.} In every data set the slowest internal mode
  separates the ring-neuron types ER3a, ER3d and ER3m from the compass
  neurons (EPG, PEN, PEG) with ER2, ER4d and ER3w. Its eigenvalue with
  all connections is 0.85 in BANC, 0.79 in the male and 0.77 in FlyWire.
  BANC with all connections sits where the male sits at 7 synapses. This
  is an unregistered observation.
\end{itemize}

\hypertarget{s8.4-measurement-or-biology}{%
\subsection*{S8.4 Measurement or
biology}}

Sources: the documentation of the BANC data release, its detector
evaluation (\url{https://doi.org/10.7910/DVN/7WTH1N}) and Bates et al.~(2026).
The registered edge list comes from the later of two detectors, with
synapses of at least 10 voxels. A second list comes from the earlier
detector, with a cutoff of 5 voxels, and is the one used in Bates et
al.~(2026). As a check, the counts between graph neurons recomputed from
the synapse table equal the registered edge list exactly (12,305,034
connections, 39,297,815 synapses).

\begin{itemize}
\item
  \textbf{BANC detects more and attaches less.} Of 255,774,906 synapses,
  77.7\% have their postsynaptic end on a segment that has not been
  identified, and 28.0\% their presynaptic end. Counting all partners a
  neuron makes 1,205 synapses, of which 22\% end on an identified
  neuron. Over 3,876 cell types present in all three data sets, the
  synapses made per neuron in BANC, counting all partners, are 1.79
  times the male (quartiles 1.26 to 2.34) and 2.45 times FlyWire (1.86
  to 3.30).
\item
  \textbf{The EB set is attached better than average.} 39\% of its
  output ends on identified neurons. In the ellipsoid-body neuropil 90\%
  of presynaptic and 51\% of postsynaptic ends are on identified
  neurons; over 24 neuropils the postsynaptic share runs from 0.17 to
  0.53.
\item
  \textbf{Excluded causes.} The detector and size cutoff: the second
  list gives stays / mixes 1.42, 1.25, 1.08, 0.81 and 0.99 at thresholds
  1 to 5, against 1.40, 1.27, 1.10, 0.87 and 0.84 for the registered
  list. Missing sections: 3 of the 1,418 sections through the set are
  sparse. Damaged regions: 5.5\% of the synapses inside the set lie in
  documented problem regions. Unattached presynaptic arbors: 78\% of the
  inputs to EB neurons come from identified neurons, against 72\% in the
  whole graph. The transmitter GABA as such: GABA types as a group are
  ordinary (Table S9).
\end{itemize}

\begin{longtable}[]{@{}
  >{\raggedright\arraybackslash}p{(\columnwidth - 6\tabcolsep) * \real{0.3583}}
  >{\raggedright\arraybackslash}p{(\columnwidth - 6\tabcolsep) * \real{0.1917}}
  >{\raggedright\arraybackslash}p{(\columnwidth - 6\tabcolsep) * \real{0.2250}}
  >{\raggedright\arraybackslash}p{(\columnwidth - 6\tabcolsep) * \real{0.2250}}@{}}
\toprule\noalign{}
\begin{minipage}[b]{\linewidth}\raggedright
Transmitter of the type (number of types)
\end{minipage} & \begin{minipage}[b]{\linewidth}\raggedright
BANC : male
\end{minipage} & \begin{minipage}[b]{\linewidth}\raggedright
BANC : FlyWire
\end{minipage} & \begin{minipage}[b]{\linewidth}\raggedright
FlyWire : male
\end{minipage} \\
\midrule\noalign{}
\endhead
\bottomrule\noalign{}
\endlastfoot
All (3,876) & 1.79 & 2.45 & 0.71 \\
Acetylcholine (2,174) & 1.85 & 2.67 & 0.67 \\
GABA (659) & 1.90 & 2.26 & 0.84 \\
Glutamate (878) & 1.67 & 2.26 & 0.74 \\
Dopamine (35) & 0.31 & 0.86 & 0.36 \\
Serotonin (7) & 0.48 & 1.93 & 0.27 \\
Octopamine (13) & 0.49 & 1.34 & 0.31 \\
\end{longtable}

\noindent\textbf{Table S9.} Synapses made per neuron, ratio between data sets over matched
cell types (median), by transmitter. BANC counted over all partners.

\begin{itemize}
\item
  \textbf{Aminergic synapses depend on the detector across the whole
  brain} (Table S9). Three partner types of the ring neurons are
  aminergic by annotation: EL, ExR2 and ExR3. EL has the 7th lowest
  ratio of the 3,876 types against the male.
\item
  \textbf{The central complex is the lowest region.} Grouping types by
  the neuropil that holds most of their output, the ellipsoid body is
  the lowest of 24 neuropils (0.76 against the male, 1.32 against
  FlyWire) and the fan-shaped body the second lowest (0.94 and 1.43).
  The published evaluation of the detector has 16 densely annotated
  cutouts, with pooled precision 0.87 and recall 0.78. Its one
  central-complex cutout, in the fan-shaped body, has recall 0.72. There
  is none in the ellipsoid body.
\item
  \textbf{It is not a uniform regional effect.} The ellipsoid-body
  neuropil holds 72\% of the synapses inside the set, and BANC has 0.56
  of the FlyWire count there. Thinning FlyWire by region only leaves
  stays / mixes at 3.08, 3.12, 3.01 and 3.06 at thresholds 1, 2, 3 and
  5.
\item
  \textbf{Inside the neuropil it is the ring neurons.} For the same
  postsynaptic cells, the ratio BANC to male is 0.18 from ring neurons
  against 0.39 from columnar neurons onto EL, 0.13 against 0.64 onto
  ExR1, 0.21 against 0.43 onto ExR2, 0.19 against 0.50 onto ExR4 and
  0.27 against 0.62 onto ExR6. Counting all partners, a ring neuron
  makes 1,123 synapses in BANC, against 1,715 in the male and 914 in
  FlyWire, where the typical cell type has 1.79 and 2.45 times the other
  animal's count.
\end{itemize}

\textbf{Not decided:} whether ring-neuron synapses are under-detected in
BANC or fewer in this animal. A densely annotated cutout in the
ellipsoid body would decide it.

\hypertarget{s9.-the-single-standing-bouts-of-the-heading-label}{%
\section*{S9. The single standing bouts of the heading
label}}

\textbf{Reading.} Fig. 4 of Seelig and Jayaraman (2015) is stored as
vector graphics in the published PDF. The markers of panels h and i were
read with their plotted coordinates and calibrated from the tick marks.
Each panel holds 445 markers; the legend prints n = 499. The 445 points
give a duration of 6.71 $\pm$ 5.14 s and a change of 0.020 $\pm$ 0.762 rad, and
panel h gives r = 0.703, which are the printed statistics. The markers
of the two panels come in the same order, and the difference computed
from panel h equals the value plotted in panel i for every bout, to
within 0.004 rad.

\begin{longtable}[]{@{}
  >{\raggedright\arraybackslash}p{(\columnwidth - 10\tabcolsep) * \real{0.1917}}
  >{\raggedright\arraybackslash}p{(\columnwidth - 10\tabcolsep) * \real{0.1333}}
  >{\raggedright\arraybackslash}p{(\columnwidth - 10\tabcolsep) * \real{0.1333}}
  >{\raggedright\arraybackslash}p{(\columnwidth - 10\tabcolsep) * \real{0.2083}}
  >{\raggedright\arraybackslash}p{(\columnwidth - 10\tabcolsep) * \real{0.1750}}
  >{\raggedright\arraybackslash}p{(\columnwidth - 10\tabcolsep) * \real{0.1583}}@{}}
\toprule\noalign{}
\begin{minipage}[b]{\linewidth}\raggedright
Bout duration (s)
\end{minipage} & \begin{minipage}[b]{\linewidth}\raggedright
Bouts
\end{minipage} & \begin{minipage}[b]{\linewidth}\raggedright
SD (rad)
\end{minipage} & \begin{minipage}[b]{\linewidth}\raggedright
Robust SD (1.4826 MAD)
\end{minipage} & \begin{minipage}[b]{\linewidth}\raggedright
Within one wedge
\end{minipage} & \begin{minipage}[b]{\linewidth}\raggedright
Beyond $\pi$/2
\end{minipage} \\
\midrule\noalign{}
\endhead
\bottomrule\noalign{}
\endlastfoot
up to 3.2 & 102 & 0.71 & 0.48 & 64\% & 5\% \\
3.2 to 4.2 & 90 & 0.62 & 0.39 & 69\% & 2\% \\
4.2 to 5.5 & 62 & 0.80 & 0.47 & 61\% & 5\% \\
5.5 to 7.5 & 63 & 0.75 & 0.44 & 62\% & 6\% \\
7.5 to 11 & 59 & 0.83 & 0.63 & 46\% & 5\% \\
11 to 36 & 69 & 0.91 & 0.71 & 42\% & 9\% \\
\end{longtable}

\noindent\textbf{Table S10.} Spread of the change in bump position by duration of the
standing bout.

\textbf{Model.} The spread grows slowly with duration (rank correlation
of the absolute change with duration 0.15, p = 0.002). Least squares on
the squared change gives an intercept of 0.43 rad\textsuperscript{2} and a diffusion
coefficient of 0.011 rad\textsuperscript{2} per second (bootstrap 95\% interval -0.0001 to
0.025). A mixture was fitted by maximum likelihood: with probability p
the position after the bout is unrelated to the one before, and
otherwise the change is normal with variance s0\textsuperscript{2} + 2Dt. The fit gives p
= 0.105 (0.060 to 0.154), s0 = 0.32 rad (0.10 to 0.44) and D = 0.012
rad\textsuperscript{2} per second (0.0049 to 0.020), so $\tau$\textsubscript{b} = 83 s (49 to
204). Without diffusion the fit is worse (likelihood ratio p = 6 $\times$
10\textsuperscript{$-$5}), without the unrelated share far worse, and an
unrelated share that grows with duration is no better (p = 0.78). The
model was chosen after looking at the points.

\textbf{Other panels.} The pooled bound of the main text, a diffusion
coefficient of at most 0.043 rad\textsuperscript{2} per second, comes from Fig. 4i. The
three further panels in darkness of the same paper (Extended Data Fig.
10b, d, f) give 0.052 to 0.076. The panel with a visible stripe has the
same spread, 0.74 rad over 4.85 s. With a stripe in view and the fly
standing the heading has a visual anchor, so most of the pooled spread
is not loss of a memory in darkness.

\textbf{Readout.} With a bump 82 to 100 degrees wide, resolving 8, 16
and 32 positions takes 2.4 to 3.6, 10 to 14 and 39 to 57 spikes in the
read window. With 10 to 13 compass neurons inside the half-maximum width
and a window of 0.1 s that is 2.3 to 2.8 Hz, 9 to 11 Hz and 37 to 45 Hz
per neuron. The one patched columnar type, P-EN, fires 3.9 $\pm$ 2.6 Hz
while the fly stands (Turner-Evans et al., 2017).

\textbf{Accuracy of the rewrite.} In the model of Turner-Evans et
al.~(2017), which has no intrinsic noise, the error between the bump and
the integrated turning grows with a diffusion coefficient of 1.82 $\times$
10\textsuperscript{$-$3} rad\textsuperscript{2} per second under a simulated turning input:
0.19 rad after 10 s and 0.47 rad after 60 s. This is an accuracy of the
rewrite and not a hold time.

\textbf{Limits.} The points were read from a figure. The unrelated share
may be jumps, fading and return of the bump, or turning during the 0.59
s after the restart over which the second estimate is averaged. All
estimates are for flies standing on a ball in darkness between walking
bouts of 2.6 to 36 s. In tethered flight in darkness the bump drifts
within seconds (Kim et al., 2017).

\section*{S10. Analyses added for this version (exploratory)}

The analyses of this section were made on 29 November 2026, after the registered run and after the review of the earlier version. They are not registered. They load the sets with the registered name rules and compute every statistic with the frozen library of the registration; the registered result file and log are unchanged. Tables S11 to S16 are written by a script directly from the result files in the folder \texttt{revision/} of the repository (release v0.4.0), so that no number is copied by hand.

\paragraph{Finite-time mixing (Table S11).} The walk conditioned to stay in each set is run from every starting neuron (or from 400 seeded starts for sets of more than 1,600 neurons), and the largest total variation distance from its stationary distribution is followed step by step. The quarter-mixing time $t_{1/4}$ replaces the relaxation time in the ratio. Henrici's departure from normality shows how far each conditioned matrix is from normal.

\paragraph{Excitatory walk (Table S12).} Only connections whose presynaptic neuron is predicted to be cholinergic are kept, and the walk is rebuilt on the giant strongly connected component of the restricted graph. Neurons without cholinergic output, among them nearly all ring neurons, leave the component.

\paragraph{Signed linear rate model (Table S13).} Activity propagates as $x(t+1)=x(t)\,\mathrm{diag}(s)P$ on the full walk. The model has no nonlinearity, so it cannot hold an activity bump; the table marks a limit of the indicator under linear dynamics.

\paragraph{Information kept by the heading label (Tables S14 and S14b).} The curves behind Table 2 of the main text, and the information actually kept at the hold times that a count of distinguishable positions would give. The count of four positions held for 7.2~s by the pooled bound corresponds to 0.95 bits, not two.

\paragraph{In silico silencing (Tables S15 and S16).} A walker that steps onto a silenced neuron is lost; the statistics of the remaining set are recomputed and compared with ten random removals of the same number of EB neurons. Table S16 gives the composition of the EB set by family and predicted transmitter in the three animals.

\setcounter{table}{10}

\setcounter{table}{10}
\begin{table}[!ht]
\centering\scriptsize\setlength{\tabcolsep}{3pt}
\caption{Finite-time mixing of the walk conditioned to stay in each set (revision analysis A1, exploratory; revision/r1\_finite\_mixing.py). $t_{1/4}$ is the quarter-mixing time from the worst starting neuron (all starts for sets up to 1,600 neurons, else 400 seeded starts); dep. is Henrici's scaled departure from normality.}\label{tab:S11}
\begin{tabular}{lrlrrrrrrrr}
\toprule
Animal & Thr. & Set & Neurons & Lost/step & $t_{\text{stay}}$ & $t_{\text{relax}}$ & Ratio & $t_{1/4}$ & $t_{\text{stay}}/t_{1/4}$ & Dep. \\
\midrule
BANC & 1 & EB & 589 & 0.133 & 9.57 & 6.86 & 1.40 & 9 & 1.06 & 0.93 \\
BANC & 1 & CX & 3,065 & 0.166 & 13.25 & 16.87 & 0.79 & 26 & 0.51 & 0.91 \\
BANC & 1 & MB & 4,898 & 0.311 & 5.80 & 16.36 & 0.35 & 28 & 0.21 & 0.89 \\
BANC & 1 & AL & 1,128 & 0.423 & 3.31 & 17.39 & 0.19 & 17 & 0.19 & 0.97 \\
\addlinespace
BANC & 5 & EB & 555 & 0.125 & 19.36 & 23.01 & 0.84 & 30 & 0.65 & 0.92 \\
BANC & 5 & CX & 2,602 & 0.192 & 21.59 & 59.55 & 0.36 & 101 & 0.21 & 0.95 \\
BANC & 5 & MB & 2,147 & 0.709 & 19.37 & 6.77 & 2.86 & 17 & 1.14 & 0.99 \\
BANC & 5 & AL & 1,068 & 0.277 & 5.61 & 14.05 & 0.40 & 22 & 0.26 & 0.98 \\
\addlinespace
Male & 1 & EB & 691 & 0.116 & 9.29 & 4.84 & 1.92 & 7 & 1.33 & 0.94 \\
Male & 1 & CX & 3,215 & 0.113 & 12.56 & 9.18 & 1.37 & 17 & 0.74 & 0.93 \\
Male & 1 & MB & 4,501 & 0.367 & 3.54 & 16.52 & 0.21 & 23 & 0.15 & 0.75 \\
Male & 1 & AL & 1,106 & 0.473 & 2.78 & 13.62 & 0.20 & 18 & 0.15 & 0.97 \\
\addlinespace
Male & 5 & EB & 691 & 0.102 & 10.20 & 6.01 & 1.70 & 9 & 1.13 & 0.94 \\
Male & 5 & CX & 3,206 & 0.107 & 14.01 & 7.71 & 1.82 & 20 & 0.70 & 0.96 \\
Male & 5 & MB & 4,498 & 0.612 & 2.45 & 5.57 & 0.44 & 10 & 0.25 & 0.97 \\
Male & 5 & AL & 1,104 & 0.424 & 3.18 & 16.48 & 0.19 & 23 & 0.14 & 0.98 \\
\addlinespace
FlyWire & 1 & EB & 683 & 0.094 & 11.97 & 4.27 & 2.80 & 7 & 1.71 & 0.96 \\
FlyWire & 1 & CX & 3,123 & 0.093 & 15.52 & 9.42 & 1.65 & 18 & 0.86 & 0.95 \\
FlyWire & 1 & MB & 5,604 & 0.413 & 3.50 & 8.71 & 0.40 & 14 & 0.25 & 0.87 \\
FlyWire & 1 & AL & 1,112 & 0.554 & 2.38 & 17.33 & 0.14 & 15 & 0.16 & 0.98 \\
\addlinespace
FlyWire & 5 & EB & 674 & 0.074 & 14.58 & 5.85 & 2.49 & 9 & 1.62 & 0.96 \\
FlyWire & 5 & CX & 3,046 & 0.078 & 19.03 & 7.81 & 2.44 & 22 & 0.87 & 0.97 \\
FlyWire & 5 & MB & 5,550 & 0.661 & 2.27 & 3.54 & 0.64 & 10 & 0.23 & 0.99 \\
FlyWire & 5 & AL & 1,111 & 0.522 & 2.61 & 22.35 & 0.12 & 20 & 0.13 & 0.98 \\
\bottomrule
\end{tabular}
\end{table}

\begin{table}[!ht]
\centering\scriptsize\setlength{\tabcolsep}{3pt}
\caption{Walk restricted to connections from neurons predicted to be cholinergic (revision analysis A2, exploratory; revision/r2r3\_signed.py). Component: neurons in the giant strongly connected component of the restricted graph.}\label{tab:S12}
\begin{tabular}{lrrlrrrrrr}
\toprule
Animal & Thr. & Component & Set & Neurons & Lost/step & $t_{\text{stay}}$ & $t_{\text{relax}}$ & Ratio & $\delta_8$ \\
\midrule
BANC & 1 & 80,289 & EB & 270 & 0.131 & 7.79 & 3.32 & 2.34 & 0.09 \\
BANC & 1 & 80,289 & CX & 1,409 & 0.185 & 65.02 & 22.64 & 2.87 & 0.90 \\
BANC & 1 & 80,289 & MB & 4,387 & 0.105 & 13.31 & 114.67 & 0.12 & 0.99 \\
BANC & 1 & 80,289 & AL & 540 & 0.372 & 3.96 & 23.78 & 0.17 & 0.77 \\
\addlinespace
BANC & 5 & 33,785 & EB & 141 & 0.088 & 11.10 & 3.87 & 2.87 & 0.14 \\
BANC & 5 & 33,785 & CX & 929 & 0.244 & 61.56 & 104.42 & 0.59 & 1.00 \\
BANC & 5 & 33,785 & MB & 584 & 0.403 & 11.59 & 6800.75 & 0.00 & 0.98 \\
BANC & 5 & 33,785 & AL & 522 & 0.178 & 11.04 & 26.53 & 0.42 & 0.86 \\
\addlinespace
Male & 1 & 103,295 & EB & 366 & 0.188 & 5.31 & 3.13 & 1.70 & 0.15 \\
Male & 1 & 103,295 & CX & 2,173 & 0.135 & 6.29 & 181.98 & 0.03 & 0.98 \\
Male & 1 & 103,295 & MB & 4,114 & 0.363 & 10.19 & 214.05 & 0.05 & 1.00 \\
Male & 1 & 103,295 & AL & 606 & 0.436 & 3.08 & 19.03 & 0.16 & 0.72 \\
\addlinespace
Male & 5 & 93,272 & EB & 311 & 0.139 & 7.02 & 3.33 & 2.11 & 0.16 \\
Male & 5 & 93,272 & CX & 2,028 & 0.111 & 8.03 & 80.12 & 0.10 & 0.98 \\
Male & 5 & 93,272 & MB & 4,106 & 0.509 & 32.49 & 39.76 & 0.82 & 1.00 \\
Male & 5 & 93,272 & AL & 605 & 0.351 & 3.63 & 22.03 & 0.16 & 0.78 \\
\addlinespace
FlyWire & 1 & 82,553 & EB & 354 & 0.161 & 6.39 & 2.65 & 2.42 & 0.09 \\
FlyWire & 1 & 82,553 & CX & 1,933 & 0.126 & 7.30 & 6.46 & 1.13 & 0.50 \\
FlyWire & 1 & 82,553 & MB & 56 & 0.959 & 1.06 & 10.28 & 0.10 & 0.76 \\
FlyWire & 1 & 82,553 & AL & 554 & 0.737 & 1.97 & 4.73 & 0.42 & 0.37 \\
\addlinespace
FlyWire & 5 & 69,280 & EB & 277 & 0.115 & 8.15 & 2.84 & 2.87 & 0.07 \\
FlyWire & 5 & 69,280 & CX & 1,631 & 0.090 & 9.23 & 11.43 & 0.81 & 0.83 \\
FlyWire & 5 & 69,280 & MB & 56 & 0.961 & 1.05 & 18.30 & 0.06 & 0.51 \\
FlyWire & 5 & 69,280 & AL & 533 & 0.720 & 2.16 & 5.30 & 0.41 & 0.40 \\
\bottomrule
\end{tabular}
\end{table}

\begin{table}[!ht]
\centering\scriptsize\setlength{\tabcolsep}{3pt}
\caption{Signed linear rate model $x(t+1)=x(t)\,\mathrm{diag}(s)P$ (revision analysis A3, exploratory; revision/r2r3\_signed.py). S1: acetylcholine $+1$; GABA, glutamate, histamine $-1$; amines and unknown 0. S2: as S1 with glutamate and unknown $+1$. Persistence $1/(1-\rho)$ and internal relaxation in steps; leading: real part of the leading eigenvalue under S1. At 5 synapses the mushroom-body set of BANC is not comparable (see main text).}\label{tab:S13}
\begin{tabular}{lrlrrrrrrr}
\toprule
Animal & Thr. & Set & S1 persist & S1 relax & S1 ratio & S1 leading & S2 persist & S2 relax & S2 ratio \\
\midrule
BANC & 1 & EB & 4.77 & 12.83 & 0.37 & -0.79 & 4.77 & 12.86 & 0.37 \\
BANC & 1 & CX & 4.78 & 12.83 & 0.37 & -0.79 & 4.77 & 19.45 & 0.25 \\
BANC & 1 & MB & 4.83 & 12.74 & 0.38 & +0.79 & 5.07 & 12.00 & 0.42 \\
BANC & 1 & AL & 2.25 & 17.58 & 0.13 & +0.55 & 2.36 & 22.23 & 0.11 \\
\addlinespace
BANC & 5 & EB & 16.29 & 28.30 & 0.58 & -0.94 & 16.28 & 28.32 & 0.57 \\
BANC & 5 & CX & 16.29 & 436.74 & 0.04 & -0.94 & 16.28 & 555.18 & 0.03 \\
BANC & 5 & MB & 19.37 & 6.77 & 2.86 & +0.95 & 19.37 & 6.77 & 2.86 \\
BANC & 5 & AL & 3.02 & 10.67 & 0.28 & +0.67 & 3.11 & 15.00 & 0.21 \\
\addlinespace
Male & 1 & EB & 2.96 & 35.43 & 0.08 & -0.66 & 2.96 & 35.18 & 0.08 \\
Male & 1 & CX & 2.96 & 35.99 & 0.08 & -0.66 & 4.64 & 8.17 & 0.57 \\
Male & 1 & MB & 2.23 & 59.58 & 0.04 & +0.55 & 2.38 & 37.44 & 0.06 \\
Male & 1 & AL & 1.96 & 30.41 & 0.06 & +0.49 & 2.08 & 27.84 & 0.07 \\
\addlinespace
Male & 5 & EB & 3.51 & 25.32 & 0.14 & -0.72 & 3.51 & 25.28 & 0.14 \\
Male & 5 & CX & 3.51 & 46.31 & 0.08 & -0.71 & 4.41 & 13.28 & 0.33 \\
Male & 5 & MB & 1.89 & 44.93 & 0.04 & +0.47 & 1.91 & 166.18 & 0.01 \\
Male & 5 & AL & 2.12 & 37.28 & 0.06 & +0.53 & 2.26 & 37.98 & 0.06 \\
\addlinespace
FlyWire & 1 & EB & 3.11 & 12.00 & 0.26 & -0.68 & 3.14 & 12.60 & 0.25 \\
FlyWire & 1 & CX & 3.11 & 12.21 & 0.25 & -0.68 & 3.57 & 18.42 & 0.19 \\
FlyWire & 1 & MB & 1.06 & 8.02 & 0.13 & -0.05 & 1.08 & 2.92 & 0.37 \\
FlyWire & 1 & AL & 1.30 & 39.16 & 0.03 & -0.23 & 1.35 & 7.76 & 0.17 \\
\addlinespace
FlyWire & 5 & EB & 4.01 & 13.53 & 0.30 & -0.75 & 4.02 & 14.38 & 0.28 \\
FlyWire & 5 & CX & 4.01 & 13.66 & 0.29 & -0.75 & 4.02 & 14.12 & 0.28 \\
FlyWire & 5 & MB & 1.08 & 19.89 & 0.05 & -0.07 & 1.10 & 4.23 & 0.26 \\
FlyWire & 5 & AL & 1.34 & 32.73 & 0.04 & -0.25 & 1.38 & 9.16 & 0.15 \\
\bottomrule
\end{tabular}
\end{table}

\begin{table}[!ht]
\centering\scriptsize\setlength{\tabcolsep}{3pt}
\caption{Information $I(t)$, in bits, that the heading label keeps about its initial value after $t$ seconds (revision analysis A4, exploratory; revision/r4\_information.py). $D$ in rad$^2$/s, $s_0$ in rad, $p$ the share of unrelated restarts. Times beyond 36~s are extrapolations.}\label{tab:S14}
\begin{tabular}{lrrrrrrrrrrrrrrr}
\toprule
Estimate & $D$ & $s_0$ & $p$ & 0.1~s & 0.5~s & 1~s & 2~s & 5~s & 7.2~s & 10~s & 20~s & 29~s & 36~s & 60~s & 103~s \\
\midrule
Pooled bound & 0.0431 & 0.000 & 0.000 & 4.03 & 2.87 & 2.37 & 1.87 & 1.21 & 0.95 & 0.71 & 0.28 & 0.12 & 0.07 & 0.01 & 0.00 \\
Fit & 0.0120 & 0.324 & 0.105 & 1.68 & 1.63 & 1.58 & 1.48 & 1.26 & 1.15 & 1.03 & 0.74 & 0.58 & 0.48 & 0.26 & 0.09 \\
Fit, $D$ low end & 0.0049 & 0.324 & 0.105 & 1.69 & 1.67 & 1.64 & 1.60 & 1.48 & 1.40 & 1.32 & 1.10 & 0.96 & 0.87 & 0.65 & 0.41 \\
Fit, $D$ high end & 0.0203 & 0.324 & 0.105 & 1.67 & 1.59 & 1.51 & 1.37 & 1.09 & 0.95 & 0.82 & 0.51 & 0.35 & 0.26 & 0.09 & 0.02 \\
\bottomrule
\end{tabular}
\end{table}

\begin{table}[!ht]
\centering\scriptsize\setlength{\tabcolsep}{3pt}
\addtocounter{table}{-1}\renewcommand{\thetable}{S\arabic{table}b}
\caption{The count-based hold time of $K$ positions, $\pi^2/(2K^2D)$, and the information actually kept at that time (revision/r4\_information.py).}\label{tab:S14b}
\begin{tabular}{lrrrr}
\toprule
Estimate & $K$ & $\log_2K$ & Hold (s) & Information kept (bits) \\
\midrule
Pooled bound & 2 & 1 & 28.6 & 0.13 \\
Pooled bound & 4 & 2 & 7.2 & 0.95 \\
Pooled bound & 8 & 3 & 1.8 & 1.95 \\
Pooled bound & 16 & 4 & 0.4 & 2.95 \\
Fit & 2 & 1 & 102.9 & 0.09 \\
Fit & 4 & 2 & 25.7 & 0.63 \\
Fit & 8 & 3 & 6.4 & 1.18 \\
Fit & 16 & 4 & 1.6 & 1.52 \\
\bottomrule
\end{tabular}
\end{table}

\begin{table}[!ht]
\centering\scriptsize\setlength{\tabcolsep}{3pt}
\caption{In silico silencing of families of the EB set (revision analysis A8, exploratory; revision/r5\_silencing.py). A walker that steps onto a silenced neuron is lost. Random: range over 10 removals of the same number of EB neurons. Visual input: TuBu, MeTu2, AOTU046 and TuTuB.}\label{tab:S15}
\begin{tabular}{lrlrrlrlrl}
\toprule
Animal & Thr. & Set & Neurons & Ratio & Random & $t_{\text{stay}}/t_{1/4}$ & Random & $\delta_8$ & Random \\
\midrule
BANC & 1 & intact & 589 & 1.40 & & 1.06 & & 0.41 & \\
 & & minus ring neurons (268) & 321 & 1.38 & 0.20 to 0.40 & 0.47 & 0.16 to 0.26 & 0.03 & 0.34 to 0.57 \\
 & & minus EPG (45) & 544 & 0.85 & 0.60 to 0.93 & 0.64 & 0.48 to 0.70 & 0.46 & 0.40 to 0.50 \\
 & & minus PEN (42) & 547 & 1.09 & 0.75 to 0.99 & 0.73 & 0.54 to 0.83 & 0.43 & 0.39 to 0.48 \\
 & & minus PEG (19) & 570 & 1.25 & 0.92 to 1.20 & 0.95 & 0.69 to 1.02 & 0.42 & 0.39 to 0.47 \\
 & & minus EPG and PEN (87) & 502 & 0.83 & 0.44 to 0.76 & 0.64 & 0.35 to 0.51 & 0.46 & 0.42 to 0.55 \\
 & & minus EL and ExR (32) & 557 & 0.49 & 0.78 to 1.01 & 0.28 & 0.57 to 0.83 & 0.77 & 0.38 to 0.50 \\
 & & minus visual input (177) & 412 & 1.36 & 0.26 to 0.46 & 1.04 & 0.19 to 0.32 & 0.38 & 0.41 to 0.59 \\
 & & minus LPsP, GLNO (6) & 583 & 1.33 & 1.24 to 1.34 & 1.01 & 0.93 to 1.10 & 0.41 & 0.39 to 0.43 \\
\addlinespace
BANC & 5 & intact & 555 & 0.84 & & 0.65 & & 0.90 & \\
 & & minus ring neurons (267) & 288 & 1.88 & 0.03 to 0.66 & 0.72 & 0.02 to 0.22 & 0.03 & 0.66 to 1.00 \\
 & & minus EPG (45) & 510 & 0.71 & 0.12 to 0.78 & 0.55 & 0.08 to 0.48 & 0.94 & 0.79 to 0.98 \\
 & & minus PEN (42) & 513 & 0.75 & 0.14 to 1.13 & 0.58 & 0.07 to 0.62 & 0.91 & 0.73 to 0.98 \\
 & & minus PEG (19) & 536 & 0.80 & 0.29 to 0.99 & 0.61 & 0.17 to 0.73 & 0.90 & 0.84 to 0.97 \\
 & & minus EPG and PEN (87) & 468 & 0.71 & 0.08 to 0.66 & 0.53 & 0.06 to 0.32 & 0.94 & 0.81 to 0.97 \\
 & & minus EL and ExR (26) & 529 & 0.58 & 0.24 to 1.18 & 0.33 & 0.17 to 0.88 & 1.00 & 0.78 to 0.95 \\
 & & minus visual input (150) & 405 & 0.85 & 0.13 to 0.88 & 0.64 & 0.08 to 0.42 & 0.89 & 0.56 to 0.96 \\
 & & minus LPsP, GLNO (6) & 549 & 0.82 & 0.48 to 0.91 & 0.62 & 0.31 to 0.69 & 0.90 & 0.88 to 0.95 \\
\addlinespace
Male & 1 & intact & 691 & 1.92 & & 1.33 & & 0.25 & \\
 & & minus ring neurons (282) & 409 & 0.33 & 0.31 to 0.63 & 0.18 & 0.21 to 0.32 & 0.55 & 0.21 to 0.40 \\
 & & minus EPG (46) & 645 & 0.71 & 0.91 to 1.40 & 0.47 & 0.60 to 0.98 & 0.38 & 0.22 to 0.34 \\
 & & minus PEN (42) & 649 & 1.22 & 1.02 to 1.46 & 0.93 & 0.65 to 0.95 & 0.30 & 0.23 to 0.33 \\
 & & minus PEG (18) & 673 & 1.60 & 1.47 to 1.85 & 1.15 & 0.94 to 1.16 & 0.27 & 0.23 to 0.29 \\
 & & minus EPG and PEN (88) & 603 & 0.70 & 0.71 to 1.06 & 0.52 & 0.44 to 0.69 & 0.37 & 0.23 to 0.37 \\
 & & minus EL and ExR (32) & 659 & 0.18 & 1.15 to 1.50 & 0.13 & 0.76 to 1.07 & 0.78 & 0.23 to 0.31 \\
 & & minus visual input (265) & 426 & 1.87 & 0.26 to 0.69 & 1.30 & 0.19 to 0.31 & 0.24 & 0.22 to 0.48 \\
 & & minus LPsP, GLNO (6) & 685 & 1.77 & 1.68 to 1.91 & 1.24 & 1.20 to 1.32 & 0.26 & 0.24 to 0.27 \\
\addlinespace
Male & 5 & intact & 691 & 1.70 & & 1.13 & & 0.36 & \\
 & & minus ring neurons (282) & 409 & 0.30 & 0.22 to 0.52 & 0.17 & 0.13 to 0.27 & 0.62 & 0.35 to 0.65 \\
 & & minus EPG (46) & 645 & 0.62 & 0.70 to 1.16 & 0.49 & 0.48 to 0.86 & 0.48 & 0.35 to 0.48 \\
 & & minus PEN (42) & 649 & 1.07 & 0.86 to 1.41 & 0.78 & 0.56 to 0.91 & 0.42 & 0.33 to 0.41 \\
 & & minus PEG (18) & 673 & 1.39 & 1.20 to 1.60 & 0.96 & 0.87 to 1.14 & 0.39 & 0.33 to 0.41 \\
 & & minus EPG and PEN (88) & 603 & 0.62 & 0.63 to 0.83 & 0.49 & 0.42 to 0.61 & 0.48 & 0.36 to 0.47 \\
 & & minus EL and ExR (32) & 659 & 0.05 & 0.95 to 1.33 & 0.04 & 0.69 to 0.96 & 0.98 & 0.33 to 0.45 \\
 & & minus visual input (265) & 426 & 1.67 & 0.14 to 0.49 & 1.12 & 0.07 to 0.32 & 0.35 & 0.30 to 0.92 \\
 & & minus LPsP, GLNO (6) & 685 & 1.54 & 1.43 to 1.64 & 1.04 & 0.98 to 1.10 & 0.38 & 0.36 to 0.38 \\
\addlinespace
FlyWire & 1 & intact & 683 & 2.80 & & 1.71 & & 0.21 & \\
 & & minus ring neurons (278) & 405 & 0.45 & 0.37 to 0.67 & 0.22 & 0.18 to 0.35 & 0.40 & 0.26 to 0.50 \\
 & & minus EPG (47) & 636 & 0.89 & 1.13 to 2.01 & 0.63 & 0.85 to 1.34 & 0.32 & 0.16 to 0.29 \\
 & & minus PEN (42) & 641 & 1.63 & 1.34 to 1.89 & 1.09 & 0.89 to 1.17 & 0.25 & 0.20 to 0.27 \\
 & & minus PEG (20) & 663 & 2.26 & 1.86 to 2.37 & 1.43 & 1.11 to 1.44 & 0.22 & 0.18 to 0.24 \\
 & & minus EPG and PEN (89) & 594 & 0.89 & 0.84 to 1.73 & 0.62 & 0.59 to 0.92 & 0.32 & 0.14 to 0.31 \\
 & & minus EL and ExR (31) & 652 & 0.17 & 1.43 to 2.44 & 0.13 & 0.96 to 1.66 & 0.80 & 0.18 to 0.26 \\
 & & minus visual input (259) & 424 & 2.72 & 0.36 to 0.69 & 1.94 & 0.23 to 0.35 & 0.20 & 0.18 to 0.39 \\
 & & minus LPsP, GLNO (6) & 677 & 2.62 & 2.28 to 2.65 & 1.61 & 1.45 to 1.63 & 0.21 & 0.20 to 0.23 \\
\addlinespace
FlyWire & 5 & intact & 674 & 2.49 & & 1.62 & & 0.38 & \\
 & & minus ring neurons (272) & 402 & 0.87 & 0.20 to 0.39 & 0.42 & 0.11 to 0.25 & 0.16 & 0.37 to 0.82 \\
 & & minus EPG (47) & 627 & 0.80 & 1.12 to 1.97 & 0.41 & 0.71 to 1.11 & 0.58 & 0.28 to 0.47 \\
 & & minus PEN (42) & 632 & 1.23 & 0.98 to 2.07 & 0.93 & 0.73 to 1.22 & 0.46 & 0.33 to 0.51 \\
 & & minus PEG (20) & 654 & 1.82 & 1.30 to 2.37 & 1.26 & 0.88 to 1.52 & 0.41 & 0.35 to 0.47 \\
 & & minus EPG and PEN (89) & 585 & 0.80 & 0.57 to 1.23 & 0.41 & 0.37 to 0.73 & 0.58 & 0.29 to 0.56 \\
 & & minus EL and ExR (31) & 643 & 0.17 & 1.21 to 1.94 & 0.10 & 0.80 to 1.22 & 0.95 & 0.30 to 0.48 \\
 & & minus visual input (256) & 418 & 2.42 & 0.20 to 0.46 & 1.58 & 0.17 to 0.31 & 0.36 & 0.33 to 0.67 \\
 & & minus LPsP, GLNO (6) & 668 & 2.25 & 1.86 to 2.53 & 1.49 & 1.25 to 1.63 & 0.39 & 0.35 to 0.42 \\
\bottomrule
\end{tabular}
\end{table}

\begin{table}[!ht]
\centering\scriptsize\setlength{\tabcolsep}{3pt}
\caption{The EB set (registered name rules) by family and predicted transmitter (revision/r5\_silencing.py; consensus prediction in the male, annotation-table prediction in FlyWire and BANC).}\label{tab:S16}
\begin{tabular}{llrrr}
\toprule
Family & Predicted transmitter & BANC & Male & FlyWire \\
\midrule
AOTU046 & glutamate & 3 & 4 & 4 \\
AOTU046 & unknown & 1 &  &  \\
EL & gaba &  &  & 17 \\
EL & octopamine & 18 & 18 &  \\
EPG & acetylcholine & 45 & 46 & 47 \\
ER & acetylcholine &  &  & 1 \\
ER & gaba & 267 & 282 & 276 \\
ER & glutamate &  &  & 1 \\
ER & unknown & 1 &  &  \\
ExR & acetylcholine & 4 & 4 & 4 \\
ExR & dopamine & 4 & 4 & 6 \\
ExR & gaba &  &  & 3 \\
ExR & glutamate & 3 & 4 & 1 \\
ExR & serotonin & 2 & 2 &  \\
ExR & unknown & 1 &  &  \\
GLNO & gaba &  &  & 3 \\
GLNO & glutamate & 4 &  & 1 \\
GLNO & unknown &  & 4 &  \\
LPsP & acetylcholine &  & 2 & 1 \\
LPsP & glutamate & 2 &  & 1 \\
MeTu2 & acetylcholine & 26 & 101 & 93 \\
MeTu2 & dopamine &  &  & 1 \\
MeTu2 & gaba &  &  & 7 \\
MeTu2 & serotonin & 1 &  &  \\
PEG & acetylcholine & 19 & 18 & 20 \\
PEN & acetylcholine & 42 & 42 & 41 \\
PEN & serotonin &  &  & 1 \\
TuBu & acetylcholine & 140 & 156 & 148 \\
TuBu & gaba &  &  & 1 \\
TuBu & glutamate &  &  & 1 \\
TuBu & unknown & 8 &  &  \\
TuTuB & acetylcholine & 1 &  &  \\
TuTuB & gaba &  &  & 2 \\
TuTuB & glutamate & 3 & 2 & 2 \\
TuTuB & unknown &  & 2 &  \\
\midrule
Total & & 595 & 691 & 683 \\
\bottomrule
\end{tabular}
\end{table}

\section*{S11. Validation report}

This report was made on 30 November 2026 for the present version. It reruns no part of the registered analysis. All files and scripts named here are in \texttt{revision/validation/} of the repository (release v0.4.0).

\paragraph{Registered files.} The confirmatory result file and run log are unchanged: their SHA-256 values, \texttt{5ffee1df97075488}\ldots{} and \texttt{da439f405715f480}\ldots, equal those in the archived releases (checked).

\paragraph{Reruns from a clean copy.} The exploratory scripts behind the numbers of the paper were run again in a fresh copy of release v0.3.0 with the folder \texttt{revision/} added and empty result folders, with the same inputs. Of 28 output files, 19 are byte-identical to the archived ones; the other 9, all logs, differ only in the printed run times. The threshold sweep of the post hoc analysis (Table 5) was not rerun, for time; its printed values were checked against the archived result files (see Numbers in the text).

\paragraph{Independent recomputation.} A second implementation, written without the frozen library (graph joined with pandas, stationary distribution from an Arnoldi eigensolver instead of power iteration, conditioned walk and Dobrushin coefficients by dense linear algebra), recomputed the registered statistics C1 to C3 and the ratios of CX, MB and AL on BANC with all connections. It is a check of the arithmetic, labelled exploratory, not a second confirmatory run. The giant component has 144,122 neurons in both. The largest relative difference over the 11 values of Table~\ref{tab:S17} is 5.9e-11. The exit channel (C4) and the reach ratio (C5) were not recomputed independently.

\begin{table}[!ht]
\centering\footnotesize
\caption{Registered statistics on BANC (all connections) against the independent recomputation.}\label{tab:S17}
\begin{tabular}{llrr}
\toprule
Set & Statistic & Registered & Independent \\
\midrule
EB & lost per step & 0.133304 & 0.133304 \\
EB & $t_{{\text{{stay}}}}/t_{{\text{{relax}}}}$ & 1.395787 & 1.395787 \\
CX & lost per step & 0.165975 & 0.165975 \\
CX & $t_{{\text{{stay}}}}/t_{{\text{{relax}}}}$ & 0.785568 & 0.785568 \\
MB & lost per step & 0.311088 & 0.311088 \\
MB & $t_{{\text{{stay}}}}/t_{{\text{{relax}}}}$ & 0.354663 & 0.354663 \\
AL & lost per step & 0.422524 & 0.422524 \\
AL & $t_{{\text{{stay}}}}/t_{{\text{{relax}}}}$ & 0.190297 & 0.190297 \\
EB & Dobrushin coefficient, 4 steps & 0.753172 & 0.753172 \\
EB & Dobrushin coefficient, 8 steps & 0.413275 & 0.413275 \\
EB & Dobrushin coefficient, 16 steps & 0.121525 & 0.121525 \\
\bottomrule
\end{tabular}
\end{table}

\paragraph{Numbers in the text.} A script formats the expected strings from the result files and searches the LaTeX source for them: Table 2 and the information figures of Sect.~3, Table 4 (from the registered result file), Table 5, all 96 entries of Table 6, the silencing figures of Sect.~8 and the count-based hold of Appendix C. 48 of 48 checks were found as printed. One rounding error of the earlier version was corrected in this way: the Dobrushin coefficient at 8 steps of the male EB set is 0.2550 and is now printed as 0.25, not 0.26.

\paragraph{References.} Each of the 51 references was looked up by its DOI, at Crossref for journal articles and books and at DataCite for arXiv and Zenodo records, and its first author, year, volume and first page compared with the entry. 49 match. Cover and Thomas carry the year 2005 at Crossref (online edition) and 2006 in print; Solms and Friston (2018) has no DOI, and its volume and pages were checked against the reference list of a later paper by the first author. Titles were compared as well, and three entries were corrected to their records. The quotations from Barron and Klein (2016) and Bates et al. (2026) were checked word for word against the full texts on 30 November 2026; the quotation from Kim et al. (2017) is carried over from the earlier version and was not checked again here.

\paragraph{What the author checked himself.} See the Methods section of the main article.

\end{document}